\documentstyle[12pt]{article}
\newcommand{\tbz}{{\mbox{\tiny            \bf             {Z}}}}

\newcommand{\bunit}{\mbox{\bf{1}}}

\newcommand{\sbtwo}{\mbox{\scriptsize            \bf            {2}}}
\newcommand{\sbthree}{\mbox{\scriptsize            \bf            {3}}}

\newcommand{\linklo}{\link^{\!\!\!\!l_0}}
\newcommand{\linkxy}{\;\link^{\!\!\!\!\!\!\!x\;\;y}}

\newcommand{\beq}{\begin{equation}}
\newcommand{\eeq}{\end{equation}}
\newcommand{\nin}{\noindent}

\newcommand{\dof}{degrees of  freedom }
\newcommand{\pcp}{partially       confining       phase }
\newcommand{\pcps}{partially        confining         phases }
\newcommand{\ppm}{$A^b_{\mu,\;Peter}, A^b_{\mu,\;Paul}, \cdots ,
A^b_{\mu,\;N_{gen.}}$ }

\newcommand{\bz}{\mbox{\bf{Z}}}
\newcommand{\bu}{\mbox{\bf{U}}}

\newcommand{\link}{\mbox{\begin{picture}(4.15,10)
\put(0,3){\circle*{2}}                     \put(0,2.75){\line(1,0){7}}
\put(7.3,3){\circle*{2}}  \end{picture}  }  }

\begin{document}
\begin{flushright}
November 19, 1993

\end{flushright}

\vspace{1.1in}

\begin{center}

{\Large \bf
Predictions  for  Nonabelian} \\
\vspace{.3cm}
{\Large \bf  Fine  Structure  Constants} \\
\vspace{.3cm}
{\Large \bf from
Multicriticality}

\vspace{50pt}

{\sl D.L. Bennett}

\vspace{6pt}

The Royal Danish School of
Pharmacy \\  Universitetsparken  2, DK-2100 Copenhagen   {\O}, Denmark \\

\vspace{18pt}

{\sl H.B.  Nielsen}

\vspace{6pt}

The  Niels  Bohr Institute \\
Blegdamsvej 17, DK-2100 Copenhagen {\O}, Denmark \\

\end{center}

\vspace{.8in}

\begin{center}
{\bf Abstract}
\end{center}


\nin In  developing  a  model  for  predicting  the  nonabelian  gauge
coupling constants we argue for the  phenomenological  validity  of  a
``principle of multiple point criticality''. This is supplemented with
the  assumption   of   an   ``(grand)   anti-unified''   gauge   group
$SMG^{N_{gen.}}\sim    U(1)^{N_{gen.}}\times    SU(2)^{N_{gen.}}\times
SU(3)^{N_{gen.}}$ that, at  the  Planck  scale,  breaks  down  to  the
diagonal subgroup. Here $N_{gen}$ is the number of  generations  which
is assumed to be 3. According to  this  ``multiple  point  criticality
principle'',  the  Planck  scale  experimental  couplings  correspond
to multiple point couplings of the bulk phase transition of a  lattice
gauge theory (with gauge  group  $SMG^{N_{gen.}}$).  Predictions  from
this principle agree  with  running  nonabelian  couplings  (after  an
extrapolation  to  the  Planck  scale  using  the  assumption   of   a
``desert'') to an accuracy of 7\%. As an explanation for the existence
of the multiple point, a speculative  model  using  a  glassy  lattice
gauge theory is presented.


\newpage

\section{Introduction}

In the present article we  describe  a  model  that  uses  bulk  phase
transition values for the plaquette action parameters at the  multiple
point in the phase diagram for a Yang-Mills lattice  gauge  theory  to
calculate the values of the nonabelian gauge  coupling  constants  for
the standard model group ($SMG$) at the Planck scale.  In  our  model,
the $SMG$ comes about as the  diagonal  subgroup  resulting  from  the
Planck scale
\footnote{The choice of the Planck scale for the  breaking
of the (grand) ``anti-unified''  gauge  group  $SMG^3$  to  its  diagonal
subgroup is not completely arbitrary insofar as gravity  may  in  some
sense be critical at the  Planck  scale.  Also,  our  predictions  are
rather insensitive to variations of up to several orders of  magnitude
in the choice of energy at which the Planck scale is fixed.}
breakdown
of our ``anti-unified'' gauge group.  This  ``anti-unified''
gauge group, on which the plaquette action of the lattice gauge  theory
is defined, is the  cartesian  product  of  a  number  $N_{gen}$  group
factors  each  of  which  is  the  standard  model
group\footnote{ The representation spectrum  of  the  standard  model
suggests\cite{mich}     using     as      the      standard      model
\underline{group}\cite{group}                     \begin{footnotesize}
\[SMG\stackrel{def.}{=}S(U(2)\times            U(3))\stackrel{def.}{=}
\left\{\left.\left( \begin{array}{cc} \bu_2 & \begin{array}{ccc} 0 & 0
& 0 \\ 0 & 0 & 0 \end{array} \\ \begin{array}{cc} 0 & 0 \\ 0 & 0 \\  0
& 0 \end{array} & \bu_3 \end{array} \right ) \right|  \begin{array}{l}
\bu_2\in   U(2),   \\   \bu_3\in   U(3),    \\    \mbox{det}\bu_2\cdot
\mbox{det}\bu_3=1 \end{array} \right\}  \]  \end  {footnotesize}  \nin
which has the same Lie algebra as  the  more  commonly  used  covering
group $U(1)\times SU(2)\times SU(3)$.} ($SMG$). The integer  $N_{gen}$
designates the number of quark and lepton generations and is taken  to
have the value 3  in  accord  with  experimental  results.  Hence  our
``anti-unified'' gauge group is

\begin{equation}   \underbrace{SMG\times   SMG\times   \cdots   \times
SMG}_{N_{gen}=3\; SMG
factors}\stackrel{def.}{=}SMG^{N_{gen}}=SMG^3. \label{et} \end{equation}

So according to our model, the physical values of the $SMG$ nonabelian
gauge couplings at  the  Planck  scale  are  equal  to  the diagonal
subgroup couplings
corresponding to the multiple point action parameters
of the Yang Mills lattice gauge theory having as the gauge
group
the ``anti-unified'' gauge group $SMG^3$.

The connection between our model and the accepted  standard  model
description of fundamental physics is made at the  Planck  scale  where
we assume that our ``anti-unified'' gauge group $SMG^3$ is broken
down to its diagonal subgroup

\begin{equation}
(SMG^3 G^3)_{diag.\;subgr.}\stackrel{def.}{=}\{(g,g,g)\mid g \in  SMG  \}
\end{equation}

\noindent which is of course isomorphic to the  $SMG$.  The  breakdown
can come about due to ambiguities  that  arise  under  group  automorphic
symmetry operations. This is usually referred to  as  the  ``confusion''
mechanism\cite{confusion,gaeta}.

In order to compare our Planck scale predictions  for  gauge
coupling constants with experiment, we of course need  to  extrapolate
experimental  values  to  the  Planck  scale.  We  do  this   with   a
renormalization group extrapolation in which a ``desert'' scenario  is
assumed in going to the Planck scale.

It should be emphasized that our model for the prediction of
gauge couplings using the ``anti-unified'' gauge  group  $SMG^3$
is  incompatible  with  the  currently  popular  $SU(5)$  or   $SU(5)$
supersymmetric grand unified models and is therefore to be regarded as
a rival to these.


Our suggestion that Nature seeks out the multiple point values
of plaquette action parameters  is  formulated  as  ``the  principle  of
multiple point criticality''. The multiple  point  (critical)  parameter
values in a phase diagram for a Yang-Mills lattice gauge theory
                     is the special point where  all  or  at  least  a
maximum number of the phases of the lattice  gauge  theory  meet.  The
multiple point critical parameter values to which we refer  are  those
that correspond to the bulk  phase  transition  of  a  (Euclideanized)
Yang-Mills lattice  gauge  theory  with  gauge  group  $SMG^3$.
Drawing upon a suggestive analogy, the idea of a multiple point  could
be likened to the triple point of water which of course is  unique  in
being the only point that is shared by the phase boundaries of all the
phases of  water  (solid,  liquid,  and  vapor)  as  depicted  in  the
parameter space spanned by temperature and pressure.

By ``phases'', we here refer to different physical behaviors that  can
be distinguished on the basis of the physics as viewed at a particular
scale. We can think of the  mean  field  approximation  ($MFA$)  which
inherently makes phase determinations on the basis of the  physics  as
seen  from  a  very  local  point   of   view   (e.g.,   the   lattice
scale)\cite{cop}. Typically, the ``phases''  (at  a  scale)  might  be
described in $MFA$ in terms such as  ``Higg's  breakdown  of  a  gauge
group to a subgroup $K$, confinement-like (strong  coupling)  behavior
w. r. t. an invariant subgroup $H \lhd K$ and Coulomb-like behavior w.
r. t. the  factor  group  $K/H$'';  the  symbol  $\lhd$  denotes  ``is
invariant subgroup of''. In this paper, we do not consider degrees  of
freedom  with  Higgs-like  behavior  (we  limit  ourselves  to  a  few
qualitative comments). This restriction amounts to  the  special  case
$SMG^{N_{gen.}}=K$  with  one  possible   phase   with
confinement-like behavior for each invariant subgroup $H$ of the  gauge
group. In  this  special  case,  the  Yang-Mills  degrees  of  freedom
corresponding to the remaining Lie subalgebra(s)  are  cosets  of  the
factor group $SMG^{N_{gen.}}/H$ and behave  as  if  in  the  ``Coulomb
phase''. Such a phase is said to be a  ``partially  confined  phase'';
i.e., ``confined'' w. r. t. the invariant subgroup $H$.


In order to ``see''  all  the  possible  \pcps  (i.e.,  one  for  each
invariant subgroup), we need to consider a class of plaquette  actions  general
enough to provoke all these possible \pcps. Such a  class  of  actions
could, for example, consist of plaquette actions that can be expressed
as a certain type of truncated  character  expansion.  An  alternative
approach corresponds to the action  ansatz  used  in  this  paper:  we
consider plaquette  actions  $S_{\Box}$  that  lead  to  distributions
$e^{S_{\Box}}$ consisting of narrow maxima centered  at  elements  $p$
belonging to certain discrete subgroups of the  center  of  the  gauge
group. The action at these peaks is then expressed as truncated Taylor
expansions around the elements $p$. With this  action  ansatz,  it  is
possible  to  provoke  confinement-like   or   Coulomb-like   behavior
\underline{independently}  (approximately  at   least)   for   the   5
``constituent'' invariant subgroups  $\bz_2,\;\bz_3,\;U(1),\;  SU(2),$
and $SU(3)$ of the $SMG$ which ``span'' the set of  ``all''  invariant
subgroups\footnote{We do not in this paper consider  the  infinity  of
invariant subgroups $\bz_N\subset U(1)$ for $N>3$.} of the $SMG$.  The
mutually uncoupled  variation  in  the  distributions  along  these  5
``constituent'' invariant subgroups is  accomplished  using  5  action
parameters: 3 parameters $\beta_1,\;\beta_2,$ and $\beta_3$ that allow
adjustment  of  peak  widths  in  the  $U(1),\;SU(2),$   and   $SU(3)$
directions \underline{and}  2  parameters  $\xi_2$  and  $\xi_3$  that
provide for adjusting the relative heights of the  peaks  centered  at
elements $p\in \mbox{span}\{\bz_2,\bz_3\}$.

The phase transitions we are interested in are first order. Hence our
model is a priori plagued by non-universality. However, our restriction
on the form of the plaquette action
nurtures the hope of at least an approximate
universality.


In order to explain why the  multiple  point  should  be  realized  in
Nature, we propose a mechanism for the stability of  this  point  that
assumes a nonlocal lattice gauge glass model with a  random  plaquette
action (e. g.; we can assume  quenched  random  values  for  character
expansion parameters). By nonlocality we  mean  that  action
terms for Wilson loops of  extent  $A$  very  large  compared  to  the
lattice constant $a$ are effectively nonlocal as seen from  scales  in
the intermediate length range $[a,A]$. We define gauge couplings  that
run due to the inclusion of successively larger Wilson loops in  going
towards the infrared. The effect of these terms on the running of  the
couplings is describable by  an  extra  term  in  the  Callan-Symanzik
$\beta$-function. The inclusion of these glassy nonlocal action  terms
in such a generalized $\beta$-function (really a multicomponent vector
of generalized $\beta$-functions) is in addition to  but  opposite  in
sign to the normal Yang-Mills renormalization group  contribution.  We
argue that rapid variations in the  generalized  $\beta$-functions  at
the multiple point can easily lead to zeros of these $\beta$-functions
close to the multiple point.
It is  estimated that already at energies near the Planck  scale  the
running plaquette action parameter values are presumably very close to
those of the ``infrared stable'' fixed point zeros of the  generalized
beta functions which in turn are close to the multiple point.

An interesting suggestion arising from our principle of multiple point
criticality is that it might offer an explanation for the smallness of
the Weinberg-Salam Higgs mass compared say to the
Planck scale because  a  principle
of multiple point  criticality,  taken  in  an  extended  form,  would
suggest that the \pcps that are accessible at the  multiple  point  by
making infinitesimal changes in the multiple point coupling values should  not
only include Coulomb-like and confinement-like  ``phases'',  but  also
Higgs-like  ``phases''.  The  sign  change   of   the   squared   mass
$m^2_{Higgs}$ of a Higgs particle at a phase boundary would imply that
$m^2_{Higgs}$ is quite small (compared to the  Planck  scale)  in  the
neighborhood of such a boundary (probably even zero if the  transition
is second order). The hierarchy problem  might  then  be  said  to  be
Nature's way of telling us that the Yang-Mills vacuum is very close to
a Higgs phase boundary!

In section 2.1 we describe the lattice gauge group  $SMG^{N_{gen}}=SMG^3$
that we use in our model. This so-called  ``anti-unified''  gauge
group consists of  a  cartesian  product  of  a  number  $N_{gen}$  of
replica of the standard model group. The  integer  $N_{gen}$  is  the
number of quark  and  lepton  generations.  The  connection  with  the
phenomenologically well established standard model group is  that  the
$SMG^3$ breaks down to the  diagonal  subgroup  of  $SMG^3$  which  is
isomorphic to the usual standard model group. In  section  2.2,  we
present some theoretical considerations that support the assumption of
the gauge group $SMG^3$.  Included  here  is  an  explanation  of  our
proposed   ``confusion\cite{confusion,gaeta}''   mechanism   for   the
breakdown of $SMG^3$ to the diagonal subgroup. Actually, any Higgs field
mechanism or  other breaking machinery will do for our  purpose  as
long as we get the  breakdown  of  the  gauge  group  $SMG^3$  to  the
diagonal subgroup $SMG_{diag}$. In section 2.3, we demonstrate  the
additivity in the inverse squared couplings of the  group  factors  in
the cartesian product $SMG^3$ in going to the diagonal subgroup.

The pivotal assumption, presented in section 3.1, is that  Nature  seeks
out values for the plaquette action parameters that correspond to  the
multiple point values in the phase diagram  of  a  Yang-Mills  lattice
gauge theory (with the  gauge  group  $SMG^{N_{gen}}=SMG^3$).  At  the
multiple point, all (or a maximal number) of the  so-called  partially
confining phases (i.e., one for each invariant  subgroup) of  the
non-simple gauge group $SMG^3$ come together. It is the multiple point
values of the plaquette action parameters for the diagonal subgroup of
the gauge group $SMG^3$ that, in the continuum limit, are equal to the
experimental gauge coupling values that have been extrapolated to  the
Planck scale. Section 3.2 is devoted to  a  short  resume  of  earlier
work. In section 3.3 we define partially confining phases. In  section
3.4 we write down a simple plaquette action  that  can  provoke  first
order phase transitions between the possible partially confining phase
combinations (we ignore some discrete subgroups of the $U(1)$ groups).

In Section 4 we evaluate the continuum limit of the  multiple
point plaquette action parameter values corresponding to the  diagonal
subgroup  of  the  gauge  group  $SMG^3$.  This  entails   essentially
corrections due to relative quantum fluctuations in the  corresponding
degrees of freedom in each  of  the  $N_{gen}$  copies  of  the  $SMG$
contained in the diagonal subgroup of $SMG^3$.


In Section 5 we present our calculated predictions  for  the
nonabelian finestructure constants
in Tables 3 and 4.

In Section 6 we outline a rather speculative model which in essence explains
the multiple point in our generalized phase diagram as a Planck scale fixed
point corresponding to zeros of the  components  of  a  multicomponent
vector of generalized $\beta$-functions.

Section  7  addresses  the  question  of  the   consistency   of   our
``anti-unified multiple point model'' with established  results.
We point out that there are only rather limit  possibilities  to  test
our model by means other than predictions of standard model parameters
such as the fine structure constants. Alternative testable predictions
of our model, which would require going far beyond the standard model,
are rather heavy cosmological strings with branches and  a  very  long
proton lifetime (if there is proton decay at all).

In Section 8, we supplement some concluding remarks with a short discussion of
the accuracy of our  numerical  predictions.  We  end  with  some
speculative comments.

\section{The anti-unified gauge group}

\subsection{The gauge group $SMG^3$}

In this paper we make two major assumptions: we assume the validity of
the principle of multiple point criticality (see next section) and  we
assume our  ``anti-unified''  gauge  group  $SMG^{N_{gen}}=SMG^3$
which breaks down at the Planck scale  to  the  diagonal  subgroup
$SMG_{diag}$. The gauge group $SMG^3$  is
described and discussed in this  section.

In the
models  we  and  others  have  developed,  the  standard  model  group
$SMG=S(U(2)\times   U(3))$   arises   as   the    diagonal    subgroup
$SMG_{diag.}\stackrel{def.}{=}\{(u,u,u)|u\in  SMG\}$  remaining  after
the   Planck   scale    breakdown    of    this    more    fundamental
(``anti-unified'')     cartesian     product     gauge      group
$(SMG)^{N_{gen.}}$ where $N_{gen.}$ is the number of quark and  lepton
families (generations).

As far as the Lie algebra is concerned, we define

\begin{small}

\begin{equation}
SMG^{N_{gen}}\stackrel{def}{=}
\underbrace{U(1)\times U(1)\times \cdots \times  U(1)}_{N_{gen}}\times
\underbrace{SU(2)\times               SU(2)\times               \cdots
\times SU(2)}_{N_{gen}}\times
\underbrace{SU(3)\times        SU(3)\times        \cdots        \times
SU(3)}_{N_{gen}}.
\end{equation}

\end{small}

\nin For the {\em group} definition of $SMG^3$, see formula (\ref{et}) and
the accompanying footnote 2 in the introduction.
We take $N_{gen}=3$ in accord with experiment. The
identification of the number of $SMG$ factors in the cartesian product
with the number of families $N_{gen}$ allows the possibility of having
different gauge quantum numbers for the $N_{gen}$ different families.

\subsection{Motivations for the $SMG^3$ gauge group}

Before  presenting  various
helpful theoretical arguments that lend credibility to the  assumption
of this Planck scale nonsimple gauge group, we point out that the most
important argument is that our ``fit'' of fine structure constants  is
so good as to justify the claim of phenomenological evidence  for  the
gauge group  $SMG^3$  as  the  relevant  one  among  the,  after  all,
countable number of candidates for more fundamental gauge groups  that
could lead to the $SMG$.

A possible theoretical
motivation for the ``anti-unified gauge  group  $SMG^3$  and  its
subsequent breakdown to the  diagonal  subgroup could start with
a scenario from ``random dynamics''\cite{randyn}: at an  energy  a  little
above the Planck scale, one has a multitude  of  gauge  symmetries
resulting from the  FNNS  exactification  \cite{fnns}  of  chance
occurrences  of  approximate  gauge  symmetries.  This  collection   of
symmetries can be expected to be  dominated  by low-dimensional groups
as such symmetries are most
likely to occur by chance. We envision that the symmetry  embodied  by
this collection of groups is broken down by a succession of  steps  the
last of which, before the Weinberg-Salam breakdown, is  the  breakdown
of the ``anti-unified'' $SMG^3$ to its diagonal  subgroup.  This
succession  of  symmetry  breakdowns  is  pictured  as  occurring   for
decreasing energies within a range of a few orders of magnitude at the
Planck scale.

The succession of breakdowns envisioned coincides  with  gauge  groups
that are more and  more  depleted  of  group  automorphisms\footnote{A
group automorphism is defined as a bijective map  of  the  group  onto
itself that preserves the group composition law. The set of all  group
automorphisms is itself a group some of  the  elements  of  which  are
inner automorphisms (i.e., just equivalence transformations within the
group).  There  can  also  be  outer  automorphisms   (essential   for
confusion) which  are  defined  as  factor  groups  of  the  group  of
automorphisms modulo inner automorphisms. }. We
(and  others)  have  proposed  a
breakdown  mechanism\cite{confusion,gaeta}  called  ``confusion''  that   is
active when gauge groups possess automorphisms.

We speculate that confusion breaking - that can be called into play by
different types of automorphisms - can successively break very general
groups with many cartesian product factors down  to  a  collection  of
groups with especially few automorphisms as is characteristic  of  the
$SMG$  itself.  It   is   noteworthy   that   the   $SMG$   has   been
shown in a certain sense \cite{skewness} to be the  group  of  rank  4
(and  dimension  less  than  19)  that  is  maximally   deficient   in
automorphisms. We propose the  group  $SMG^3$  as  the  last
intermediate step on the way to the $SMG$.

\subsubsection{Gauge group breakdown by confusion}

We now briefly explain
how the confusion breakdown mechanism functions for  gauge  groups
with outer automorphisms. First it is argued that,  in  the  spirit  of
assuming a fundamental physics that can be taken as  random,
one is forced to allow for the possibility of having  quenched  random
``confusion surfaces'' in spacetime. The defining  property  of  these
surfaces  is  that  (e.g.  gauge)  fields  obey  modified   continuity
conditions at such surfaces; for example, the permutation of  a  gauge
field with an automorphic image of the field can  occur.  A  nonsimply
connected  spacetime  topology  is  essential  for  the  presence   of
nontrivial confusion surfaces; a discrete spacetime structure such  as
a lattice is inherently nonsimply connected because of  the  ``holes''
in the structure.

The essential feature of  the  ``confusion''  breakdown  mechanism  is
that,  in  the  presence  of  ``confusion  surfaces'',  the   distinct
identities of a field and its automorphic image can be maintained {\em
locally} but not {\em globally}. To see  how  this  ambiguity  arises,
imagine taking a journey along a  closed  path  on  the  lattice  that
crosses a confusion surface at which  the  labels  of  a  gauge  group
element and its automorphic image are permuted. Even if one could,  at
the start of the journey, unambiguously assign say the names ``$Peter$''
and ``$Paul$'' to two gauge  fields  related  by  an  automorphism,  our
careful accounting of the field identities as  we  travel  around  the
loop would not, upon arriving back at the starting point,  necessarily
be in agreement with the  names  assigned  when  we  departed  on  our
journey. So an attempt to make independent global gauge transformations of
$Peter$ and $Paul$ (sub)groups would not succeed.
Therefore, for the action at confusion surfaces, there is not invariance
under global gauge transformations of the whole gauge group but only under
transformations of the subgroup left invariant by the automorphism.

The ambiguity under the automorphism caused by confusion (surfaces)
is removed by the breakdown of the gauge group
to the maximal subgroup which is left invariant under the  automorphism.
The diagonal  subgroup of the cartesian
product of isomorphic groups is
the maximal invariant subgroup  of  the  permutation  automorphism(s);
i.e., because the diagonal subgroup is the subgroup left invariant  by
the automorphism, it has  the  symmetry  under  gauge  transformations
generated by constant gauge functions  (corresponding  to  the  global
part  of  a  local  gauge  transformation)  that  survives  after  the
ambiguity caused by the automorphism is removed by  breakdown  to  the
diagonal subgroup.

For the purpose  of  illustrating  a  possible  origin  of  the
``anti-unified'' gauge group  and  its  subsequent  breakdown  to  the
standard model group, we describe  two  important  examples  of  group
automorphisms - {\em examples 1 and 2} below - that call the confusion
mechanism into play:

\begin{description}

\item[{\em  Example  1.}]Many  groups  have  a   charge   conjugation-like
automorphism corresponding in the $SU(N)$ case to complex  conjugation
of the matrices element by element. While for $SU(2)$ this is an inner
automorphism, it is for higher $SU(N)$ groups an  outer  automorphism.
According to the speculated confusion mechanism, such a  group  should
break down to the  subgroup  consisting  of  only  the  real  matrices
which  is  the  largest  subgroup  that   is   invariant   under   the
automorphism.  If  the  group  is  provided  with  C-breaking   chiral
fermions, the automorphism can be broken in this way thereby thwarting
the ``attack'' from the confusion mechanism.


\item[{\em Example 2.}] There can be automorphisms under the permutation
of identical group factors in a cartesian product group: we argue that
the  symmetry   reduction   (at   the   Planck   scale)   from
$(SMG)^{N_{gen.}}$ to $SMG=S(U(2) \times U(3))$ is  triggered  by  the
symmetry under the automorphism  that  permutes  the  $N_{gen}$  $SMG$
factors in $(SMG)^{N_{gen.}}$.

\end{description}

Elaborating briefly on {\em example 1} above, we point  out  that  with
the exception of the semisimple groups such as $SU(2)$, $SO(3)$ ,  the
odd $N$ spin or $SO(N)$-groups and the symplectic groups,  all  groups
have  outer  automorphisms  of  the  complex  conjugation-  or  charge
conjugation-like type. Following  a  series  of  confusion  breakdowns
activated by charge conjugation-like automorphisms, we expect that the
(intermediate)   surviving   gauge    symmetry    (i.e.,    that    of
$(SMG)^{N_{gen.}}$)  must  have  matter  fields  that   break   charge
conjugation-like symmetries. In other  words,  the  presence  of  such
matter fields serves to protect the surviving  symmetry  from  further
breakdown by eliminating the possibility for further confusion  of  the
surviving group with its automorphic image under charge conjugation.

In particular, we expect that a necessary condition for  the  survival
of gauge groups like $U(1)$ and $SU(3)$ is the presence of
some  matter  fields  not  invariant  under  charge  conjugation.
Protection against this sort of breakdown can be  provided  by  chiral
fields that break the charge conjugation symmetry of the gauge fields.
In the case of the Standard Model,  left-  and  right-handed  fermions
always appear in different representations so that confusion breakdown
by way of a charge conjugation automorphism is not possible. In  fact,
the number of particles in a single generation in combination with the
rather intricate way these are represented in the Standard  Model  can
be shown to be the simplest possible manner in which  gauge  anomalies
can be avoided\cite{anomalies,korfu}.

As mentioned in {\em example  2}  above,  we  assume  that  the  final
breakdown of gauge symmetry by confusion  (at  the  Planck  scale)  is
activated by the automorphism that permutes the  $N_{gen}$  isomorphic
cartesian product factors $SMG$ in $(SMG)^{N_{gen.}}$. The elimination  of
the ambiguities that  can  arise  in  trying  to  keep  track  of  the
identities of a group element and its  automorphic  image  under  such
permutations coincides with the breakdown to the standard model  group
$SMG=S(U(2) \times  U(3))$  which,  being  the  diagonal  subgroup  of
$(SMG)^{N_{gen.}}$, is invariant under the automorphism that  permutes
the $SMG$ group factors in $(SMG)^{N_{gen.}}$. In order for this final
confusion breakdown to work effectively, the cartesian product factors  of
$(SMG)^{N_{gen.}}$ must presumably be truly isomorphic  -  i.e.,  the
matter field content of each factor must  essentially  have  the  same
structure. This  combined  with  the  fact  that  one  usual  fermion
generation is known to provide the least  complicated  arrangement  of
particles that avoids gauge anomalies would strongly suggest that  the
$N_{gen}$ factors of $(SMG)^{N_{gen.}}$ are simply dull repetitions of
the Standard Model Group each one of which can have  its  own generation.
Each  of  the  $N_{gen}$  factors   is   the
``ancestor'' to one of  the  $N_{gen}$  generations  of  the  diagonal
subgroup identified with the usual Standard Model Group.

It should be emphasized that all  the  confusion  breakdowns  -  those
utilizing   a   series    of    charge    conjugation    automorphisms
leading to $(SMG)^{N_{gen.}}$ as well as the final confusion breakdown
of the $SMG^3$  to  the  diagonal  subgroup  that  is  caused  by  the
permutation  automorphism
                  - are assumed to take place within a  rather  narrow
range of energies at the Planck scale.

Before leaving the confusion breakdown mechanism, we should point  out
that any mechanism that  breaks  the  $SMG^3$  down  to  the  diagonal
subgroup would suffice for our model. A Higgs  field  mechanism  could
for example provide an  alternative  to  the  confusion  mechanism  of
breakdown.

\vspace{.8cm}

We end the discussion of the motivation for the  ``anti-unified''
gauge  group  $SMG^{N_{gen}}$  with  some  remarks  on   some   recent
work\cite{gap} of a slightly different nature.  In  this  work  it  is
argued that the hierarchical mass spectrum of  quarks  and  leptons  -
especially the large mass gaps between generations  -  calls  for  the
association  of  different  gauge   quantum   numbers   to   different
generations. The feature that appears to be required is that, for example,
the (gauge) quantum number difference between left and right $\mu$  is
not the same as that between the left and  right  $e$.  This  suggests
gauge group extensions (at the fundamental scale) having many features
in common with $SMG^3$ although the large top quark  mass  presents  a
problem for the gauge group $SMG^3$.

\subsection{Additivity of the inverse squared couplings}

The breakdown of the  group  $SMG^3$  to  the  diagonal  subgroup  has
consequences for the gauge couplings that we now briefly  describe.
Recalling that the diagonal  subgroup of $SMG^3$ corresponds  by  definition
to identical excitations of the $N_{gen}=3$ isomorphic gauge
fields (with the gauge couplings absorbed)
and using the names $Peter$, $Paul,\cdots $  as  indices  that
label the $N_{gen.}$ different isomorphic cartesian product factors of
$(SMG)^{N_{gen.} }$, one has\footnote{As it is $gA_{\mu}$ rather
than $A_{\mu}$ that appears in
the (group valued) link variables $u\propto e^{iagA_{\mu}}$, it is the
quantities $(gA_{\mu})_{Peter}$, $(gA_{\mu})_{Paul}$, etc. which are equal in
the diagonal subgroup.}


\begin{equation} (g{\cal A}_{\mu}(x))_{Peter} =(g{\cal A}_{\mu}(x))_{Paul} =
\cdots  =   (g{\cal   A}_{\mu}(x))_{N_{gen.}}   \stackrel{def.}{=}(   g{\cal
A}_{\mu}(x))_{diag.};\label{diagconfig} \end{equation}

\nin this has the consequence that the common $(gF_{\mu \nu})^2_{diag}$
in each term of the lagrangian density
for $(SMG)^{N_{gen.}}$ can be factored out:

\begin{equation}  {  \cal
L}=-1/(4g_{Peter}^2)(gF_{\mu                      \nu}^a(x))^2_{Peter}
-1/(4g_{Paul}^2)(gF_{\mu      \nu}^a(x))^2_{Paul}       -       \cdots
-1/(4g_{N_{gen.}}^2)(gF_{\mu \nu}^a(x))^2_{N_{gen.}} \end{equation}

\begin{equation} =(-1/(4g_{Peter}^2)  -1/(4g_{Paul}^2)-\cdots
-1/(4g_{N_{gen.}}^2)) \cdot (F_{\mu \nu}^a(x))^2_{diag}  =-1/(4g_{diag}^2)
\cdot (gF_{\mu\nu}^a(x))^2_{diag}. \end{equation}

The inverse squared couplings for the
diagonal subgroup is  the sum of the inverse  squared  couplings
for each of the $N_{gen.}$ isomorphic  cartesian  product  factors  of
$(SMG)^3$. Additivity in the inverse squared couplings in going to the
diagonal subgroup  applies  separately  for  each  of  the invariant
Lie
subgroups\footnote{For $U(1)$, a modification is  required.}
$i \in \{SU(2), SU(3)\}\subset SMG$. However, for $U(1)$ it  is  possible
to    have    terms    in    the    lagrangian     of     the     type
$F_{\mu\nu}^{Peter}F^{\mu\nu\;Paul}$  in  a   gauge   invariant   way.
Therefore it becomes more complicated as to how one should  generalize
this notion of additivity. Terms of this type can  directly  influence
the $U(1)^3$  continuum  couplings\footnote{In  seeking  the  multiple
point for $SMG^3$, one is lead to seek criticality separately for  the
cartesian  product  factors  as  far  as  the  nonabelian  groups  are
concerned. For $U(1)$, recent progress suggests that one  should  seek
the multiple point for the whole group $U(1)^3$ rather than  for  each
of the $N_{gen.}=3$ factors $U(1)$ separately.  The  reason  for  this
complication concerning abelian groups  (continuous  or  discrete)  is
that these have subgroups and thereby invariant subgroups  (infinitely
many for continuous abelian groups) that cannot be regarded as being a
subgroup of one of the $N_{gen.}$ factors of $SMG^3$  or  a  cartesian
product of such subgroups. Work on this problem  is  in  progress  and
appears at this time to lead to a phenomenologically desirable  factor
``6''                  for                  the                  ratio
$\frac{\alpha_{crit.,\;U(1)}}{\alpha_1(\mu_{_{Pl.}})}$          (where
$\alpha_{crit.,\;U(1)}$ is the critical coupling for the  gauge  group
$U(1)$) instead of the  factor  ``3''  (from  $N_{gen.}$)  that  would
naively be expected for this ratio by analogy to the  predictions  for
the nonabelian couplings.}. But for the nonabelian couplings we simply
get

\begin{equation}
\frac{1}{g_{i,diag}^2}=\frac{1}{g_{i,Peter}^2}+\frac{1}{g_{i,Paul}^2}+\cdots
+\frac{1}{g_{i,N_{gen}}^2}     \;\;(i\in     \{SU(2),\;SU(3)\})      .
\label{confadd} \end{equation}

Assuming that the inverse squared
couplings for a given $i$ but different labels  $\{
Peter,$ $ Paul, \cdots,N_{gen.} \}$ are all driven to the  multiple
point in accord with the principle of multiple point criticality (see next
section),
these $\{ Peter, Paul, \cdots,N_{gen.} \}$ couplings
all become equal to the multiple point value $g_{i,multi.\; point}$; i.e.,:

\begin{equation}
\frac{1}{g_{i,Peter}^2}=\frac{1}{g_{i,Paul}^2}=\cdots
=\frac{1}{g_{i,N_{gen}}^2}=\frac{1}{g_{i,\;multi.\; point }^2}.
\end{equation}

We see that the inverse squared coupling $1/g_{i, \;  diag}^2$  for  the
$i$th subgroup of the diagonal subgroup is enhanced  by  a
factor $N_{gen}$ relative to the corresponding subgroup $i$ of each of
the $N_{gen}$ cartesian product  factors  $Peter$,  $Paul,
\cdots, N_{gen.}$ of $(SMG)^{N_{gen.}}$:

\begin{equation}
\frac{1}{g_{i,diag}^2}=\frac{N_{gen}}{g_{i,\;multi.\; point }^2}.
\end{equation}

It is this weakening of the coupling for each of the  subgroups  $i\in
\{SU(2),SU(3)\}$ of the  diagonal  subgroup  (i.e.,  the  $SMG$)  that
constitutes the main role of the anti-unification scheme in our model.
Anticipating the discussion of the role of the multiple point  in  the
next  section,  we  point  out  prematurely  that  while  it  is   the
$g_{i,\;multi.\; point }$ (i.e., $i=SU(2)$ or $SU(3)$) which are to be
identified with  the  critical  values  (at  the  multiple  point)  of
coupling  constants  for  the  bulk  phase  transition  of  a  lattice
Yang-Mills theory with gauge group $i$, it is the
$g_{i,\;diag.}= g_{i,\;multi.\; point }/\sqrt{N_{gen.}}$ that, in  the
continuum  limit,  are  to  be  identified  with   the   corresponding
experimentally  observed  couplings   extrapolated   to   the   Planck
scale\cite {kim,amaldi}.

\vspace{.8cm}

In  summary,  section  2.1 describes the
``anti-unified'' gauge group $SMG^{N_{gen}}=SMG^3$. Section  2.2
is  devoted  to  an  attempt  to  motivate  the  assumption   of   the
``anti-unified'' gauge group and includes a description  of  the
``confusion'' mechanism of breakdown of this  group  to  the  diagonal
subgroup (which is isomorphic to the standard  model  group
$SMG$).
Other breakdown mechanisms for the gauge group $SMG^3$ would  also  do
as long as they lead to the diagonal subgroup.

Assuming that Nature seeks out parameter values at the  point
in the phase diagram of a Yang-Mills lattice gauge theory (with  gauge
group $SMG^3$) where a maximum number of phases come  together  (i.e.,
the multiple point; see next section), it is seen in section 2.3  that
the breakdown to the  diagonal  subgroup  of  $SMG^3$  results  in  an
$N_{gen}$-fold decrease in the critical values  of  the  finestructure
constants corresponding to the bulk phase  transition  values  at  the
multiple point of $SMG^3$.

\section{The principle of multiple point criticality}

\subsection{Statement of principle}

The main point of the present work is to put forward the idea which we
state as  a  principle  -  the  principle  of  multiple point criticality:

\hspace{.2cm}

\nin {\em At the fundamental scale (taken to be the Planck scale), the
actual running gauge coupling constants correspond to the  multiple point
critical values in the phase diagram of a lattice gauge theory.}

\hspace{.2cm}

\nin The multiple point is a point in the phase  diagram  of  the
lattice gauge theory at which all - or at least many - \pcps\footnote
{Partially confining phases are explained in section 3.3}  meet.  This
point corresponds to  critical  values  for  the  parameters  used  to
describe the form of the  action.  In  the  rather  crude  mean  field
approximation that we consider, there is one \pcp\footnote{In reality,
such  phases  are  not  necessarily  separated  by  a  phase  boundary
everywhere in the action parameter space; e.g., phase boundaries  that
end at a critical point can be circumvented in going from one phase to
another.} for each invariant subgroup of the gauge group  -  including
discrete invariant subgroups.

In section 3.2 we comment briefly on related earlier work. This
is  followed  by  a  definition  of  partially  confining  phases   in
section 3.3. Section 3.4 is used to explain how we choose enough
parameters in the plaquette action so as to: 1) have the existence  of
the \pcps we want to consider and 2) be able  to  make  these  phases
come together at a point - the multiple point - in the space of  these
parameters.

\subsection{Relation to earlier work}

In earlier  work,  critical  values  for  gauge  couplings  were
calculated using an incomplete set of critical action parameters.
We used this restricted form of criticality\footnote{In earlier  work,
we claimed that couplings should be critical (in the incomplete sense)
at the fundamental scale (presumably the Planck scale)  using  several
types of arguments: 1) couplings  exceeding  critical  values  at  the
fundamental scale would lead to confinement ``already'' at the  Planck
scale and thereby leave no  trace  of  the  corresponding  degrees  of
freedom at lower energies; 2) for some reason - perhaps as  protection
from ``Higgsning'' - Nature seeks out the strongest allowed couplings;
we referred to this tendency as ``saturation'' of the inequality to be
fulfilled in order to avoid the Planck scale confinement mentioned  in
1).} of gauge couplings to  fit  the  number  of  generations\cite{ngen}
which, at the time, was not known. The idea behind the  principle  of
multiple point criticality is to calculate  critical  gauge  couplings
from a set of (multiple point)  critical  action  parameters  that  is
complete in the sense that  a  confinement  phase  for  any  invariant
subgroup of the gauge group can be obtained for  some  choice  of  the
parameters of this  set.  This  has  several  advantages  compared  to
earlier work.

1) A weak point  in  our  previous  procedure
for calculating ``critical'' couplings using  an  incomplete  set  of
critical action parameters is that, when applied to  the
groups $SU(2)$ and $SU(3)$ (the groups  to  which  we  have  primarily
applied this procedure), these phase transitions strictly speaking  do
not exist with the simplest type of action\footnote{In previous  work,
these phase transitions were ``found'' only in the context of our very
crude $MFA$ approximation.}. There is no singularity in  the  partition
function - in particular, not in a corrected mean field approximation  - but
rather only a peak or kink  indicating  behavior  that  presumably  is
qualitatively similar to a phase transition. With the  multiple  point
criticality principle, we avoid this  embarrassing  situation  because
the multiple point for the groups $SU(2)$ and $SU(3)$ corresponds to a
proper phase transition although it is possible  to  circumvent  this
transition by going around the critical point at  the  terminus  of  a
phase boundary. That it is possible to go continuously from one phase  to  the
other  is  due  to  having  extra  characters  in  the  action  (e.g.,
corresponding to the adjoint representation).

2) If we restrict ourselves to simply searching for  parameter  values
that yield critical gauge couplings  for  the nonsimple  gauge  group
$SMG^3$, it is necessary to  supplement  this  restricted
``criticality
principle'' with a statement that it is to be applied separately  to
all $N_{gen}=3$ isomorphic  $SU(2)$  and $N_{gen.}=3$ isomorphic  $SU(3)$
invariant subgroup factors of $SMG^3$.
This is because the prediction of the three  continuum  gauge
couplings  from  our  model  requires  that  each  of  the  $N_{gen.}$
cartesian product factors $SU(2)$ as well as  each  of  the  $N_{gen}$
factors $SU(3)$ have critical coupling values. However,  according  to
the principle of  multiple  point  criticality,  critical  values  are
sought out for all invariant subgroups (including isomorphic copies of
the same invariant subgroup and discrete invariant subgroups).

3) We think that almost any mechanism that explains why Natures  seeks
parameter values corresponding to a  phase  boundary  for  some  group
would almost certainly be  applicable  to  invariant  subgroups;  this
would amount to the prediction that Nature seeks out  a  phase
boundary point such that infinitesimal changes in  the  parameters  of
the theory would  give  access  to  the  \pcps  corresponding  to  all
invariant subgroups of the group in question.  Assuming  that  such  a
point exists, it is, by definition, just the multiple point.


In summary, we have pointed out in  this  section  that,  from  the
standpoint  of  earlier  work,  the  postulate   of   multiple   point
criticality has several advantages.

\subsection{Classification of phases}

Since we postulate that Nature seeks a special point in plaquette action
parameter space - the multiple point - where many phases come together, we
need now to develope an idea about what phases exist. This is the purpose of
this subsection.

First it should be made clear that when we talk about phases of
nonabelian groups at the Planck scale,   we mean
{\em first order} phase transitions between ``phases'' that are
lattice artifacts.
All these ``phases'' will, for
sufficiently long distances, turn out to be confining with no long range
correlations (corresponding to finite glueball masses)  when, as is the
case in this section,  matter fields are ignored.

So for us, the interesting phases are separated by {\em  first  order}
phase transitions. Such phases  are  governed  by  which  {\em  micro}
physical fluctuation patterns yield the maximum  value  of  $\log  Z$.
Qualitatively  different  short  distance  physics  could  consist  of
different distributions of group elements along various  subgroups  or
invariant subgroups of the gauge group for different regions of (bare)
plaquette action parameter space. It is therefore the physics  at  the
scale of the lattice that is of interest because it is  lattice  scale
physics that dominates the different $\log Z$  ansatzes  that  prevail
(i.e., are maximum) in different parameter space regions separated  by
first order transitions. However,  this  does  not  mean  that  longer
distance behavior is  unchanged  in  passing  from  one  ``phase''  to
another. As an example, consider the string tension at the  transition
between two different {\em lattice  scale  phases}  that  both  really
correspond to  confining  phases  in  the  usual  sense:  in  what  we
designate as confining in the  mean  field  approximation  ($MFA$)  or
``confinement at the lattice scale'', the string tension has an  order
of magnitude given by dimensional arguments. On  the  other  hand,  in
what we call the ``Coulomb  phase  at  the  lattice  scale'',  or  the
Coulomb phase in the $MFA$ approximation, the string tension  is  much
smaller; i.e,. smaller by an exponential factor.

Since we want to relate these lattice artifact first order phase transitions
to experimental observations, we are obliged to  take the point of view
that these artifacts have - one way or another - physical reality.

Now we want to  assign  names  to  the different lattice artifact phases, i.e.,
qualitatively  different  physical
behaviors of the vacuum  of a lattice gauge theory at the lattice scale.
We use a classification according  to  whether  or
not there is spontaneous breakdown of  the  gauge  symmetry  remaining
after an incomplete choice of gauge  that  we  here  take  to  be  the
(latticized) Lorentz gauge.  We  therefore  consider   the   transformation
properties of  the  vacuum  under  gauge  transformations  that  leave
the Lorentz
gauge condition (i.e., $\prod_{\;\link\;
\begin{tiny} \mbox{emanating from } \end{tiny} \bullet  }  U(\link)=1$
for all sites \begin{Huge}$\cdot$\end{Huge}) intact.

Examples of such gauge transformations are those that can be generated
either by a constant gauge coordinate function $\alpha^a(x)=\alpha^a$ or
a linear gauge coordinate function $\alpha^a(x)=\alpha^a_{\mu}x^{\mu}$
corresponding  to  gauge  transformations  designated  respectively  as
$\Lambda_{Const.}(x)=    e^{i\alpha^at^a}$    and     $\Lambda_{Linear}(x)=
e^{i\alpha^a_{\mu}t^ax^{\mu}}.   $   The   gauge   coordinate    function
$\alpha^a(x)$ is the $a$th  ``color''  coordinate  of  a  Lie  algebra
vector $\alpha^a(x)t^a$. Here $t^a$ denotes an element
of the Lie algebra satisfying the commutation relations
$[t^a,t^b]=c^{ab}_ct^c$ where the $c^{ab}_c$ are the structure
constants.


On the basis of the transformation  properties  of  the  vacuum  under
gauge   transformations    of    the    types    $\Lambda_{Const.}(x)=
e^{i\alpha^at^a}$                                                  and
$\Lambda_{Linear}(x)=e^{i\alpha^a_{\mu}t^ax^{\mu}}$,   we   want    to
classify    the    different     possible     ``\pcps''     of     the
vacuum\cite{frogniel}.  In  our  classification,  there  can  be   one
``\pcp'' for each possible  combination  of  a  subgroup  $K$  and  an
invariant subgroup $H$ such that $K \subset SMG^3$ and $H \lhd K$. The
subgroup $K$, which we refer to  as  the  ``unHiggsed''  subgroup,  is
defined   as   the   group   of   constant    gauge    transformations
$\Lambda_{Const.}$  that  leaves  the  vacuum   invariant\footnote{The
vacuum invariance referred to  really  means  the  invariance  of  the
coefficients $\langle D^{(\mu)}_{ij}(U(\linkxy)) \rangle_{vac}$ in  an
expansion in  (matrix  elements  of)  continuous  unitary  irreducible
representations $D^{(\mu)}_{ij}(U(\linkxy))$. The  expansion  referred
to is that corresponding to some  link  variable  probability  density
function $P(U(\link^{\!\!\!\!l_0})) = \int \prod_{\link  \;\;  \not  =
\;\; \linklo} d^{{\cal H}aar}U(\link) e^S.$} . The \dof  corresponding
to the homogeneous space $SMG^3/K$ are  accordingly  ``Higgsed''.  The
invariant subgroup $H$, referred  to  as  the  confined  subgroup,  is
defined  as   the   invariant   subgroup   $H\lhd   K$   of   elements
$h=\exp\{i\alpha^1_at_a\}$ such that  the  gauge  transformations  with
linear gauge function $\Lambda_{Linear}$ exemplified
by\footnote{In  the  quantity
$x^1/a$, $a$ denotes the lattice constant; modulo  lattice  artifacts,
rotational invariance allows the (arbitrary) choice of  $x^1$  as  the
axis     $x^{\mu}$     that     we     use.}$\Lambda_{Linear}\stackrel
{def.}{=}h^{x^1/a}$ leave the vacuum invariant.

For  the  \dof
associated with the factor group $K/H$, there  is  invariance  of  the
vacuum  expectation  value  under  the  corresponding  constant  gauge
transformations, while there is spontaneous breakdown under the linear
gauge transformations corresponding to these \dof. Such \dof  will  be
said to demonstrate ``Coulomb-like'' behavior.

The  use  of  the  adjectives   ``Higgs-like'',   ``Coulomb-like''   and
``confinement-like''  in  classifying     qualitatively   different
physical behaviors at  a  given  scale  (here  the  Planck  scale)  is
motivated from results obtained using  the  mean  field  approximation
$(MFA)$ which intrinsically distinguishes phases on the basis  of  the
qualitative differences in the physics  that are  discernible  at  the
scale of the lattice constant.

Assuming as we do here that all vacuum Yang-Mills \dof of the  $SMG^3$
are ``unHiggsed'' (i.e., {\em all}  $\Lambda_{Const.}\in  SMG^3$  leave
the vacuum invariant), we  can  then  define the  Yang-Mills  \dof
corresponding to an invariant  subgroup  $H  \lhd  SMG^3$  as  being
``confined'' if $H$ is the  maximal  subgroup  of  $SMG^3$  such  that
linear gauge transformations $\Lambda_{Linear}\in H$ leave the  vacuum
invariant. The remaining Yang-Mills
\dof take values in the set of cosets belonging to  the  factor  group
$SMG^3/H$ and behave as if in a ``Coulomb phase''. We  call  this  the
``\pcp'' that is confining w.r.t. $H$.
In seeking the multiple point, we  seek  the
point or surface in parameter  space  where  ``all'' (or many)
partially  confining
phases ``touch'' one another.

\subsection{Generalized action is required}

We have seen that there are many  possible  phases  according  to  the
classification of the  foregoing  section.  If  we  want  to  let  the
principle of multiple point criticality guide us  in  calculating  the
values of gauge couplings found in Nature, we need to find  a  set  of
action parameters that distinguishes all these phases from each  other
and also allows all these phases to come together at a  point  in  the
phase diagram. This of course requires the inclusion of  a  sufficient
number of suitable parameters in the ansatz for the  plaquette  action
used. This is what we seek to set up in this section:  a  sufficiently
extensive class of parametrized plaquette actions.

To find out which action  parameters  to  use,  we  have  developed  a
technique for the construction of crude phase diagrams for gauge
groups such as $SMG$, $SMG^3$ and $U(N)$ (collectively referred to  by
the symbol ``$G$''); $U(N)$ is useful because it has many features  in
common the $SMG$ and $SMG^3$ while,  for  expositive  purposes,  being
simpler to deal  with.  As  these  groups  are  nonsimple,  there  are
nontrivial partially confining phases corresponding  to  ``confining''
behavior for a number of invariant subgroups that can include discrete
invariant subgroups. All these groups  have  $U(1)$  as  an  invariant
subgroup. However, we do not in the present work  consider  the  \pcps
corresponding to all the discrete (invariant) subgroups of the abelian
$U(1)$ subgroup;  rather  we  restrict  our  considerations  to  those
discrete  invariant  subgroups  of  $U(1)$  which  are  also  discrete
invariant subgroups of nonabelian Lie subgroup factors (i.e.,  $\bz_2$
and $\bz_3$ for $SMG$ or $\bz_N$ for $U(N)$). A  consequence  of  this
restriction is that we are dealing with a finite number $n$ of \pcps.

Having a multiple point for $n$ such phases
presupposes  an  appropriate  choice  of  the  action
parameter space: for the parameter set chosen, it must be possible  to
bring all $n$ phases together at a point.  That  we  want  {\em all} $n$
phases to meet  at  the  (multiple)point  further  requires  that  the
parameter set choice can separate all $n$ phases (at  least  near  the
multiple point). This means that we need a parameterization  that  can
make the various invariant subgroups go  more  or  less  independently
from a confinement-like to a Coulomb-like  behavior.

The most general form for the action requires an  infinite  number  of
parameters that  could,  in  principle  at  least,  be  taken  as  the
coefficients of an expansion in group characters corresponding to each
imaginable way of associating group characters with Wilson loops.  For
such a very general action, the infinite dimensional phase diagram for
the theory with a gauge group having  $n$  invariant  subgroups  would
have $n$ phases that meet  along  an  infinite-dimensional  ``multiple
point '' critical surface of codimension $n-1$. The boundaries between
the various phases - if  first  order  -  would  be  characterized  by
singularities in the  first  derivative  of  $\log  Z=\mbox{max}\{\log
Z_L|L\lhd G\}$. At the boundary  delineating  a  phase  for  which  an
invariant subgroup $H_i$ is realized as being confinement-like,  $\log
Z_{H_i}$ would dominate in a  part  of  the  neighborhood  of  such  a
boundary. We define the quantity $\log Z_H$ as being $\log Z$ when the
latter is calculated as if we have a field configuration  distribution
for which the lattice gauge theory is in a  phase  for  which  $H$  is
``confining'' and  $G/H$  is  ``Coulomb-like''\footnote{i.e.,  Bianchi
identities are ignored for \dof along $H$ and $MFA$ is used w.  r.  t.
$G/H$ with link averages $\langle U(\link)_{G/H}\rangle \neq 0$.}.

In practice, we use an  ansatz  action  having  a  finite,  manageable
number $d$ of $\beta$-parameters that span a $d$-dimensional  subspace
of the most general (infinite dimensional) parameter  space.  In  this
subspace, it is often possible to  find  a  set  of  parameter  values
$\vec{\beta}_{crit.}\stackrel{def.}{=}\{\beta_{1,\;crit.}       \cdots
\beta_{d,\;crit.}\}$ lying in the infinite dimensional multiple  point
critical surface of codimension $n-1$ that is  embedded  in  the  most
general parameter space as described in the preceeding  paragraph.  In
the $d$-dimensional subspace, such a  $\vec{\beta}_{crit.}$  coincides
with the multiple point or, more generally,  lies  in  a  manifold  of
multiple points that formally can be defined as follows:

\beq \{\vec{\beta}|  \forall \;  H_i   \lhd   G \;  \exists   \;\;
\vec{\epsilon}_{infinitesimal}=(\epsilon_1, \cdots , \epsilon_d)  \;\;
[\log   Z_{H_i}(\vec{\beta}+\vec{\epsilon})=\log    Z=\mbox{max}\{\log
Z_K|K\lhd G\}]\}.\eeq

If $d=n-1$ for the $d$-dimensional subspace of the  infinite  dimensional
action parameter  space,  multiple points  are  referred  to  as
generic multiple points and can be found systematically for  many
parameter choices. If $d>n-1$, then we would generally expect to  have
a    multiple    point     critical     surface\footnote{Our     crude
approximation\cite{long} used to extract qualitative features  of  the
phase diagram implies that the continuum couplings at different points
on the multiple surface  have  the  same  values.  Variations  in  the
continuum couplings along the multiple point surface cannot be seen in
the approximation where we  use  the  truncated  Taylor  expansion  of
(\ref{tay2}) and (\ref{tay3}).}  of  codimension  $n-1$.  If  $d<n-1$,
there can be what we call nongeneric  multiple  points  but  only  for
judicious choices of the $d$-dimensional parameterization.

In order to see the difference between a  generic  and  a  non-generic
multiple point, let us consider a journey in  action  parameter  space
that starts at  a  random  point  and  subsequently  seeks  out  phase
boundaries for which a  successively  greater  number  of  phases  are
accessible by making infinitesimal changes in the action parameters at
the points along the journey. In the generic case, the codimension $codim$ of
the boundary  goes  up  by  one  each  time  an  additional  phase  is
encountered so that a phase boundary surface/curve is in contact  with
$codim+1$ phases. In  the  non-generic  case,  the  number  of  phases
accessible goes up faster than the codimension - at least once in  the
course of the journey in parameter space. So for the non-generic case,
it is possible to have points  in  phase  space  (e.g.,  the  multiple
point) forming a surface at which infinitesimal variations  in  action
parameters give access to a number of phases exceeding $codim+1$ where
$codim$ is the codimension of this surface.

\subsubsection{The modified Manton action}

We find nongeneric multiple points using an ansatz that restricts  the
class of actions to those that lead to distributions $e^{S_{\Box}}$ of
plaquette variables consisting of narrow\footnote{We are assuming that
a procedure using a weak coupling approximation is  valid  even  in  a
neighborhood at the critical $\beta$'s.  The  approximation  in  which
$\frac{1}{6}\beta$ is considered to be very small is  discussed  below
and in reference \cite{long}.} ``peaks'' centered at the elements  $p$
belonging to discrete subgroup(s) of the center of  the  gauge  group.
For expositional purposes, we can think of the gauge group $U(N)$  and
the discrete invariant subgroup $\bz_N \subset U(N)$. Formally we  can
characterize the simplest of this restricted class of actions  as  the
projection of the class of all possible actions onto the  subspace  of
actions having the form of second order  Taylor  expansions  in  group
variables with one Taylor expansion  at  each  element  $p\in  \bz_N.$
Assuming for simplicity that the peaks  expanded  around  each  element
$p\in \bz_N$ are symmetric (so that odd-order derivatives  vanish),
the simplest of this restricted class of plaquette actions (neglecting
zero order terms) would then be of the form

\beq  \left.  \frac{\partial^2   S_{act.}}{\partial   \omega_a\partial
\omega_b}\right|_                           {\vec{\omega}^{(p)}}
(\omega_a-\omega^{(p)}_a)(\omega_b-\omega^{(p)}_b) \label{tay2}\eeq

Here the $\omega_a$    are  coordinates  on  the  group
manifold in a neighborhood of an element $p\in \bz_N$. The coordinates
at                  $p$                  are                   denoted
$\vec{\omega}^{(p)}=(\omega_1^{(p)},\omega_2^{(p)},\cdots            ,
\omega_{dim(U(N))}^{(p)})$.

 From the assumption  of  a
distribution $e^{S(\Box)}$ of \underline{narrow} peaks centered at the
elements $p \in \bz_N$, a group element  not  close  to  some  $p  \in
\bz_N$ leads to a vanishing value of $e^{S(\Box)}$. This means that  a
nonvanishing value of $e^{S(\Box)}$ at  a  given  group  element
$\vec{\omega}=(\omega_1, \omega_2, \cdots ,\omega_{dim(U(N))})$
gets its value in our ansatz action solely from the
Taylor expansion centered at just one element $p \in \bz_N$ (i.e., the
one  for  which  the  quantity   $\sum_{a,b}g^{ab}(\vec{\omega}^{(p)})
(\omega_a-\omega^{(p)}_a)(\omega_b-\omega^{(p)}_b)$ is minimum).  Here
$g^{ab}$ is  the  metric  tensor defined by requiring
invariance under left and right group multiplication supplemented with
normalization conventions.  We  define  the
quantities $\beta_i$ (at the point $p$) by

\beq  \left.  \frac{\partial^2   S_{act.}}{\partial   \omega_a\partial
\omega_b}\right|_                                  {\vec{\omega}^
{(\vec{\omega}^{(p)})}}
\stackrel{def.\;of\;\beta_i}{=}\sum_i\beta_ig_i^{ab}
(p) \label{tay3} \eeq

\nin where the  index $i$ labels the Lie subgroups
of the gauge group invariant w.r.t. the algebra.
For the group $U(N)$,  $i\in  \{U(1),SU(N)\}$
and we take $p\in \bz_N$. So the  action  we  are
considering - from now on we call it  the  modified  Manton  action  -
leads to Gaussian peaks at each $p$; $\beta_i$ is the action parameter
that determines the width of the Gaussian distribution along the $i$th
Lie subgroup. We assume that $\beta_i$ is the same\footnote{In a  more
sophisticated ansatz, this need not be assumed.} at all elements $p\in
\bz_N$. A  full  specification  of  our  modified
Manton action also  requires  parameters  that  specify  the  relative
height    of    peaks    at    the    different     elements     $p\in
\bz_N$. For $N=2$ or $3$, one parameter - denote it by  $\xi_N$  -  is
sufficient.

Using the parameters of the modified Manton action (e.g., for  $U(N)$,
the parameters are $\beta_i$ and $\xi_N$ described above),  nongeneric
multiple points are readily found. The selection of subgroups to which
these parameters correspond make up what we call  the  ``constituent''
invariant subgroups\footnote{ By constituent  invariant  subgroups  we
refer essentially to the cartesian product  factors  of  the  covering
group together with a selection  of  the  discrete  subgroups  of  the
center - namely the ones of special importance in obtaining the  gauge
group as a factor group (e.g., $\bz_2$ is  of  special  importance  in
obtaining the factor group $U(2)$ from the covering group  $U(1)\times
SU(2)$   because   $U(2)\overline{\simeq}(U(1)\times   SU(2))/\bz_2$.}
(e.g., for $U(N)$, the ``constituent invariant subgroups  are  $U(1)$,
$SU(N)$, and $\bz_N$). The corresponding set  of  parameters  has  the
advantage that they are essentially independent: variation of  one  of
these parameters leads to a change in the width  of  the  distribution
$e^{S_{\Box}}$  along  the  corresponding  ``constituent''   invariant
subgroup that is approximately uncoupled from the distributions of group
elements along other ``constituent'' invariant  subgroups.  Also,  all
possible invariant subgroups $H$ of the gauge groups  considered  here
can be constructed as factor groups associated with a  subset  of  the
set of the ``constituent'' invariant subgroups.

For the $SMG$, the ``constituent''  invariant  subgroups  are  $U(1)$,
$SU(2)$,  $SU(3)$,  $\bz_2$,  and  $\bz_3$.   The   $\beta_i$   ($i\in
\{U(1),SU(2),SU(3)\}$)  constitute  three   of   the   five   required
parameters  for  the  modified  Manton  action.  The   remaining   two
parameters we shall  designate  as  $\xi_2$  and  $\xi_3$.  These  are
associated with the discrete invariant subgroups $\bz_2$  and  $\bz_3$
and determine the  relative  heights  of  peaks  of  the  distribution
$e^{S_{\Box}}$ at various elements  $p\in  \mbox{span}\{\bz_2,\bz_3\}$.

The  modified
Manton action for the $SMG$ is

\beq   S_{\Box}(U(\Box))=    \left\{    \begin{array}{l}    \sum_{i\in
\{U(1),SU(2),SU(3)\}}\beta_idist_i^2(U(\Box),      p)+\log      \xi(p)
\;\;\;\mbox{for $U(\Box)$ near $p\in \mbox{span}\{\bz_2,\bz_3\}$ }  \\
\approx  -\infty\;\;\;  \mbox{for  $U(\Box)$  not   near   any   $p\in
\mbox{span}\{\bz_2,\bz_3\}$} \end{array} \right. \label{modman} \eeq

\nin where the  symbol  $dist_i^2(U(\Box),p)$  denotes  the  component
of the distance\footnote{Left-right invariance of a Riemann space metric
$ds^2=g_{ab}d\omega^a d\omega^b$ specifies $dist_i^2$ for each simple
invariant subalgebra $i$ up to a normalization factor. The decomposition
of $dist(p,u)=\int^u_p ds$ into components $dist_i$ along the $i$th invariant
subgroup is at least well defined
for small distances.} between the group element $U(\Box)$  and  the  nearest
element $p\in \mbox{span}\{\bz_2,\bz_3\}$ along  the  $i$th  invariant
Lie subgroup and

\beq  \log  \xi(p)=\left  \{  \begin{array}{ll}  0  &  \mbox{for   }
p\in  \bz_3  \\   \log   \xi_2   &   \mbox{for   }
p\not \in \bz_3 \end{array} \right \} + \left \{ \begin{array}{ll} 0 &
\mbox{for } p\in \bz_2 \\ \log \xi_3 &  \mbox{for
} p\not \in \bz_2 \end{array} \right \}. \eeq

\nin This action gives rise to  a
distribution  $e^{S_{\Box}}$  having   6   Gaussian   peaks   at   the
elements\footnote{The notation used here is that of  the  footnote  in
the beginning of the Introduction where the $SMG$ is embedded into $SU(5)$.}

\beq p_r\stackrel{def.}{=}
\left( \begin{array}{cc} e^{\frac{i2\pi r}{2}}\bunit^{2\times 2} &
\begin{array}{ccc} 0 & 0  &  0
\\ 0 & 0 & 0 \end{array} \\ \begin{array}{cc} 0 & 0 \\ 0 & 0 \\ 0 &  0
\end{array} & e^{\frac{-i2\pi r}{3}}\bunit^{3\times 3} \end{array} \right )
\in \mbox{span}\{\bz_2,\bz_3\}=\bz_6 \;\;\;\mbox{with  } r=0,1,\cdots,5.  \eeq

\nin all  having
   widths     $(2\beta_1)^{-1/2}$,     $(2\beta_2)^{-1/2}$,     and
$(2\beta_3)^{-1/2}$   in
respectively the $U(1)$, $SU(2)$, and $SU(3)$ subgroup directions. For
$r=0$ (i.e., at the group identity), the peak height is 1; for  $r=3$,
(i.e., at the nontrivial  element  of  $\bz_2$)  the  peak  height  is
$\xi_2$; for $r=2,3$ (i.e., the nontrivial elements of  $\bz_3$),  the
peak heights are $\xi_3$; for $r=1,5$, the  peak  heights  are  $\xi_2
\xi_3$.

Note that the assumption of very sharp peaks at the  elements  $p  \in
\mbox{span}\{\bz_2,\bz_3\}$  means  that   actions   of   this   class
are in essence completely specified by parameters corresponding to the
coefficients of second order terms in their  Taylor  expansions  about
these elements $p$. Higher order  terms  in  the  Taylor  expansion  are
irrelevant. Roughly speaking, this also means that physical quantities
such as continuum couplings and the  log$Z_H$  for  various  invariant
subgroups $H$ (and therefore the realizable phases) can only depend on
the coefficients of second order action  terms.  Hence,  the  multiple
point critical surface can be expected to be approximately parallel to
surfaces of constant continuum coupling  values  thereby  yielding  an
approximate universality.

\subsubsection{Factorizing the free energy}

The critical coupling values for the transitions on the  lattice
which we are interested in here happen to be so relatively weak that a
weak coupling approximation is not meaningless. We therefore make  use
of  such  an  approximation  in  conjunction   with   a   mean   field
approximation ($MFA$) in our very crude  exploratory  studies  of  the
qualitative features to be expected for the phase diagram of a lattice
gauge theory. A sensible weak  coupling  approximation  requires  that
$\beta^{-1}$ is small compared to the squared extension of  the  group
where of course $\beta$ is the coefficient of the  real  part  of  the
trace of a plaquette variable in the action. Because  we  assume  weak
coupling, we use the approximation of a  flat  space  measure  in  the
evaluation of functional integrals  with  the  limits  of  integration
extended to $\pm \infty$.

It is natural to enquire as to how such a weak coupling  approximation
can have a chance of being sensible in dealing with  the  onset  of  a
confinement-like phase at the lattice phase transition.  Recall  first
that we use as the defining feature of a confinement-like  phase  that
Bianchi identities can be ignored  to  a  good  approximation\footnote{This is
consistent with the definition in section 3.3: when fluctuations are so strong
that gauge symmetry is not broken by a gauge transformation with a linear
gauge function $\Lambda_{Linear}$ (leading to a translation of the
gauge potential $A_{\mu}$  by a constant), then the
fluctuations can also be assumed to be so strong
that the effect of Bianchi identities are washed out.}.  But  the
variables to which the Bianchi identities  apply\footnote{Recall  that
Bianchi identities  impose  a  constraint  (e.g.,  modulo  $2\pi$  for
$U(1)$) on the divergence of flux from a volume enclosed by
plaquettes} are not plaquette variables but rather the  variables
taken by  3-dimensional  volumes  enclosed  by  plaquettes  -  in  the
simplest case, just the cubes  bounded  by  six  plaquettes.  Now  the
distribution of such cube variables is the 6-fold convolution  of  the
distribution  of  plaquette  variables were it not
for Bianchi identities.  So  if  the  distribution  of
plaquette variables has a width proportional to  $\beta^{-1}$,
the width of the distribution of group elements taken by  cubes
enclosed  by  six  plaquettes  is  proportional  to  $(\beta/6)^{-1}$.
Therefore, the question of the  validity  of  using  a  weak  coupling
approximation at the  phase  transition  on  a  lattice  is  really  a
question of whether the number 6 can be regarded as large compared  to
unity. Accepting  this  as  true,  we  can  conclude  that  even  when
$\beta^{-1}$ is small compared to the squared extension of  the  group
as required for a meaningful weak coupling approximation, the quantity
$(\beta/6)^{-1}$ is large enough compared to the squared extension  of
the  group  so  as  to  justify  the  the  use  of  the  Haar  measure
distribution for the Bianchi-relevant cube variables obtained  as  the
convolution of six plaquette variable distributions. This  amounts  to
ignoring the Bianchi identities. A phase for which Bianchi  identities
can be ignored is, according to our ansatz, a confinement-like phase.

In this approximation we obtain an  expression  for  the  free  energy
$\log        Z$         in         terms         of         quantities
$Vol(L)\stackrel{def.}{=}\frac{vol(L)}{``fluctuation\;vol.''}\cdot(\pi e)^
{\frac{dim(L)}{2}}$  where $L$  is
a factor group/invariant subgroup of the $SMG$
and   $vol(L)$
is the volume of $L$. The ``fluctuation volume'' is defined as the width of
the flat distribution that yields the same entropy as the original
distribution; i.e.,

\[ \Delta S_{ent}= < -\log(e^{\beta dist^2})-\log(``flat\; distribution'')>=0.
\]

\nin For large $\beta$ (weak coupling approximation), we have the
approximation $Vol(L)\approx \beta^{\frac{dim(L)}{2}}vol(L)$ which we shall
use in the sequel.

For the \pcp that is confined w. r. t.  the  invariant
subgroup $H$, the free energy per active\footnote{Active link means  a
link not fixed by a  gauge  choice  in  say  the  axial  gauge.}  link
is\cite{long}                  \begin{equation}                   \log
Z_{per\;active\;link}=\mbox{max}_H\{\log Z_H|H\lhd G\} \eeq

\nin where

\beq \log Z_H =                    \log
[\frac{(\pi/6)^{(dim(G/H))/2}}{Vol(G/H)}                       ]+2\log
[\frac{(\pi)^{(dim(H))/2}}{Vol(H)}] \label{fe} \end{equation}

\[= \log  [\frac{(\pi/6)^{(dim(G))/2}}{Vol(G)}
]+\log      [\frac{(6\pi)^{(dim(H))/2}}{Vol(H)}]
\]

\nin where it is understood that  $\log  Z_H$  is  calculated  using  an
ansatz that results in confinement for the invariant subgroup $H$  and
Coulomb-like behavior for the  factor  group  $SMG/H$.

In our approximation (\ref{fe}), it can be shown that at the  multiple
point, any two invariant subgroups $H_1$ and $H_2$ of the gauge  group
must satisfy the condition

\beq \log(\sqrt{6\pi})^{dim(H_J)-dim(H_I)}=\log\frac{Vol(H_J)}{Vol(H_I)}
\label{piv}\eeq

\nin where it is recalled that the quantity $Vol(H_J)$ is  essentially
the volume of the subgroup $H_J$ measured in  units  of  the  critical
fluctuation volume.
In special case where $H_I=\bunit$, we get

\beq \frac{\log Vol(H_J)}{dim(H_J)}=\log(\sqrt{6\pi});\eeq

\nin i.e., for {\em any} invariant subgroup $H_J$  the  quantity  $Vol
(H_J)$ per Lie algebra dimension must be equal to  the  same  constant
(``$\sqrt{6\pi}$'') at the multiple point.

This factorization property can be realized for all the  13  invariant
subgroups $H \lhd SMG$ (or all the 5 invariant subgroups $H\lhd U(N)$)
using the previously defined set of constituent  invariant  subgroups.
That is, it is possible to factorize $Vol(H)$ into a product  of  some
subset of a common set of 5 factors  $(1/p_i)Vol(K_i)$  (with  $p_i\in
N^+$) (3 such factors for $U(N)$)  corresponding  to  the  constituent
invariant subgroups $K_i \in  \{\bz_2,\bz_3,U(1),SU(2),SU(3)  \}$  for
the $SMG$ (for $U(N)$, the constituent invariant subgroups are $K_i\in
\{SU(N),U(1),\bz_N \}$). Then it is  possible  by  adjustment  of  the
parameters   $\beta_1,\beta_2,\beta_3,\xi_2,\xi_3$   to    make    the
quantities $\log Z_{H\;per\;active\;link}$ equal  for  each  invariant
subgroup $H\lhd SMG$ (the same  applies  to  each  invariant  subgroup
$H\lhd U(N)$ using the parameters  $\beta_1,\beta_N,\xi_N$).  This  is
equivalent to finding  a  {\em  nongeneric}  multiple  point  (because
$5<n_{SMG}-1=13-1$ for the $SMG$ and $3<n_{U(N)}-1=5-1$ for $U(N)$).


We explicitly demonstrate the factorizability of $Vol(H)$ in the sense
that we show that it is of  the  form  $Vol(H)=\mbox{product  of  some
factors} (1/p_i)Vol(K_i)$ for each  invariant  subgroup  $H$  of  both
$SMG$ and $U(N)$. To do this, we use a calculational trick in which we
replace each $H$ by a cartesian product group  related  to  $H$  by  a
homomorphism  that  is  locally  bijective.  This  (to  $H$)   locally
isomorphic cartesian  product  group  consists  of  the  covering  Lie
(sub)groups  corresponding  to  the  gauge  \dof  that  the  invariant
subgroup  $H$  involves  supplemented  by  the  discrete   constituent
invariant subgroups contained in these  Lie  subgroups.  For  all  the
invariant subgroups $H$, the cartesian product group  replacement  can
be obtained by simply omitting factors in the cartesian product  group
replacement $\bz_2\times \bz_3\times U(1)\times SU(2)\times SU(3)$ for
the whole $SMG$. Of course such a cartesian product group  in  general
differs in global structure from the invariant subgroup  $H$  that  it
replaces. However, as we are only interested in the quantity  $Vol(H)$
for invariant subgroups $H$, we can use a  correction  factor\footnote
{We define the quantity $p_{_{H}}=\#(H\cap D)$ where  $D$  (which  has
$\#(D)=36$) is the discrete  subgroup  of  the  center  that  must  be
divided out of the cartesian product  group  $\bz_2\times  \bz_3\times
U(1)\times SU(2)\times  SU(3)$  in  order  to  get  the  $SMG$;  i.e.,
$((\bz_2\times   \bz_3\times    U(1)\times    SU(2)\times    SU(3))/D)
\overline{\simeq}SMG  \stackrel{def.}{=}  S(U(2)\times   U(3))$.}
$1/p_{_{H}}$ to adjust the quantity $Vol$ of the cartesian  product
group replacement for $H$ so as to make it equal to $Vol(H)$.

As an example, consider the  invariant  subgroup  $H=U(2)\subset  SMG$
which is locally isomorphic to the cartesian product group $U(1)\times
SU(2)\times \bz_2\times \bz_3$. By this we  mean  that,  assuming  the
modified  Manton   action   (\ref{modman})   and   a   weak   coupling
approximation, the cartesian  product  group  $U(1)\times  SU(2)\times
\bz_2 \times \bz_3$ simulates the subgroup $U(2) \subset SMG$  in  the
sense that the regions on the group manifold of $U(2)\subset  SMG$  in
which the probability distribution $e^{S_{\Box}}$ is concentrated  can
be  brought  into  a  one  to  one  correspondence  with  centers   of
fluctuation sharply peaked around  points  in  the  cartesian  product
group.  In  other  words,  for  $U(2)\subset  SMG$,   the   region   of
correspondence with the cartesian product group is the composite of  6
small      neighborhoods      around      the      elements      $p\in
\mbox{span}\{\bz_2,\bz_3\}$. Even though the cartesian  product  group
in this example contains $2\cdot 2\cdot 3$ elements for  each  element
in $U(2)$, the action on the cartesian product group is defined so  as
to be $-\infty$ everywhere except at one  of  the  these  12  elements
where  this  action  then  has  the  same  value  as  the  action   at
corresponding element of $U(2)$.

In order to make the quantity $Vol(U(1)\times SU(2)\times  \bz_2\times
\bz_3)$ equal to $Vol(U(2))$ (for $U(2)\subset SMG$) , the former must
be reduced by a factor $p_{U(2)}$ obtained as follows:  Remember  that
the $U(1)$ embedded in the $SMG$ has a length $6\cdot  2\pi$  so  that
the  $U(2)$  subgroup  lying  in  the  $SMG$  is  $(U(1)_{12\pi}\times
SU(2))/\bz_2$.      Comparing      the      quantity       $Vol(U(2))=
Vol(U(1)_{12\pi})\cdot Vol(SU(2))/\bz_2$ and the  quantity  $Vol$  for
the locally isomorphic  cartesian  product  group:  $Vol(U(1)_{12\pi})
\times Vol(SU(2))\times  Vol(\bz_2)\times  Vol(\bz_3))$,  it  is  seen
that, relative to $Vol$ for the cartesian product group, the  quantity
$Vol(U(2))$ is down by $(\#\bz_2)\cdot (\#\bz_2)\cdot (\#\bz_3)=2\cdot
2\cdot 3=12\stackrel{def}{=}p_{U(2)}$.

In Table~1, we  demonstrate  explicitly  that  the  volume  correction
factors $1/p_H$ for all the invariant subgroups  $H\lhd  SMG$  can  be
factored into a subset of 5 factors $1/p_i$ associated  with  each  of
the     ``constituent''     invariant     subgroups      $K_i      \in
\{\bz_2,\bz_3,U(1),SU(2),SU(3)\}$. For a given $i$,  $p_i$  is  always
the same in any $p_H$ in which $p_i$ is a factor. Listed in the  first
column of Table~1 are the quantities $Vol(H)$  for  all  13  invariant
subgroups $H$ of the $SMG$;  listed  in  the  second  column  are  the
quantities $Vol$ for the corresponding, locally  isomorphic  cartesian
product groups. The third column consists  of  the  volume  correction
factors $1/p_H$ by  which  the  quantities  $Vol$  for  the  cartesian
product group in the second column must be multiplied in order
to get the corresponding quantity $Vol(H)$ in the first column. In the
next five columns, we give the factorization of the correction factors
$1/p_H$ into subsets of five rational quantities $1/p_i$ with  $i  \in
\{\bz_2,\bz_3,U(1),SU(2),SU(3)\}$ that are associated  with  the  five
quantities $Vol(\bz_2)$, $Vol(\bz_3)$, $Vol(U(1))$, $Vol(SU(2))$,  and
$Vol(SU(3))$. Table 2 is constructed in an analogous fashion for the 5
invariant subgroups of $U(N)$ using $\bz_N,\; U(1)$,  and  $SU(N)$  as
the constituent invariant subgroups. For both the  $SMG$  and  $U(N)$,
the important point is that,  for  any  invariant  subgroup  $H$,  the
factorization

\beq Vol(H)=\prod_i(\frac{Vol(K_i)}{p_i}) \;\;\;  (\mbox{\footnotesize
$i$ runs over a  subset of constituent  invariant  subgroups})
\label{volk} \eeq

\nin is such that the correction factor $1/p_i$
corresponding to a  given  constituent  invariant  subgroup  $K_i$  is
always  the  same  (unless  the  quantity  $Vol(K_i)$  is  absent   in
the product (\ref{volk}) in which  case  there  is  no  entry  in  the
column headed by $1/p_i$) in Tables~1~and~2.

The meeting of 13 \pcps at the (nongeneric) multiple point in
the phase diagram for the $SMG$ in the 5-dimensional action  parameter
space is virtually impossible to depict clearly in a figure.  However,
the group $U(N)$, which has many features in common with the $SMG$,  has
a phase diagram with a nongeneric multiple point in 3  dimensions
when we use an action ansatz analogous to that  used  for  the  $SMG$:
Gaussian peaks at elements of $\bz_N\subset U(N)$.
The  phase  diagram for $U(2)$ seen in Figures~1~and~2  shows,
in our approximation, the
5 \pcps (corresponding to
the 5 invariant subgroups of a $U(N)$ group)
that meet at the multiple point.

\subsubsection{Need for discrete subgroup parameters}

It is instructive to answer the question: Why do we need the  discrete
group parameters? Recall that in our modified Manton action, there can
be  sharp  ``peaks''  in  the  distribution $e^{S_{\Box}}$ of  plaquette
variables
 centered at nontrivial elements of discrete  subgroups.
However, to motivate the answer to  our  question,  we  revert  for  a
moment  to  an   action   $S_{\Box}$   leading   to   a   distribution
$e^{S_{\Box}}$ of plaquette variables with just one ``peak''  (at  the
identity) - this is just the normal Manton action.  Then  the quantities $Vol$
corresponding to the {\em the same  Lie  algebra  ideals}  in  the  Lie
Algebra of the $SMG$ - i.e., volumes measured in units proportional  to
the
fluctuation volume - obey the relations

\beq \begin{array}{lcl}
Vol(U(1)_{subgr.}) & = & 6Vol(SMG/(SU(2)\times SU(3))) \\
Vol(SU(2)_{subgr.}) & = & 2Vol(SMG/U(3)) \\
Vol(SU(3)_{subgr.}) & = & 3Vol(SMG/U(2)) \\
Vol(U(2)_{subgr.}) & = & 3Vol(SMG/SU(3)) \\
Vol(U(3)_{subgr.}) & = & 2Vol(SMG/SU(2)) \\
Vol(SU(2)\times SU(3))_{subgr.} & = & 6Vol(SMG/U(1)).
\end{array} \label{vols} \eeq

The important feature of this list is that it relates  the  volumes  of
subgroups and factor groups that correspond to the  same  Lie  algebra
ideals (in the Lie algebra of the $SMG$). It is seen that  the  volume
of a subgroup with a given Lie algebra is larger than  a  factor
group with the same Lie algebra by an  integer  factor  equal  to  the
number of center elements of the subgroup that are identified  in  the
factor group.

Without the extra ``peaks'' of the modified Manton action, it  is  not
possible to vary the volume on the left hand side of  one  of  the
equations (\ref{vols})  {\em independently} of the volume on  the
right hand side. Because of the (nontrivial) integer factor  disparity
in the volumes on the two sides of the equations  (\ref{vols}),  it  is
not   possible   to   have,   for   example,   $Vol(U(2)_{subgr.})   =
Vol(SMG/SU(3))$ at the same point in the $\beta$ parameter  space.  In
particular, two such volumes can never have critical  values  for  the
same values of the $\beta$'s which means that  the  two  corresponding
partially  confining  phases  (i,e.,  confinement  w.r.t.  $U(2)$  and
$SU(3)$) cannot meet at a multiple point.

This feature is seen in Fig. 2 which shows  the  phase  diagram
for   the   gauge   group   $U(2)$.   In   the   plane   defined    by
log$Vol(\bz_2)=\log 2$ (the maximum value of log$Vol(\bz_2)$), we have
the phase diagram corresponding  to  the  normal  Manton  action  (one
``peak'' in the  distribution  of  plaquette  variables,  centered  at
the     identity).     Due     to     the      fact      that      say
$Vol(SU(2)_{subgr.})=2Vol(U(2)/U(1))$ (measured in the  same  unit  of
volume which is  the   fluctuation  volume  which  in  turn  is
proportional                      to                      $\prod_{i\in
\{U(1),SU(2),SU(3)\}}(2\beta_i)^{-dim(i)/2}$), it  is  impossible
to have $Vol(SU(2)_{subgr.})=Vol(U(2)/U(1))$ for the  same  values  of
the $\beta$ parameters; i.e., because the volumes of the subgroup  and
factor group corresponding to the same Lie algebra differ by a  factor
two, these two volumes cannot be critical for the same set of  $\beta$
parameters. This in turn precludes phases  partially  confined  w.r.t.
$SU(2)$   and   $U(1)$   from   coming   together.   In   the    plane
log$Vol(\bz_2)=\log 2$ of Fig. 2, it is indeed seen that  these  \pcps
do not touch; the maximum number of phases that come together in  this
plane is three (i.e., not all four possible  phases)  where  three  is
generic number of phases that can meet in two dimensions.

In    order     to     succeed     in     having,     for     example,
$Vol(SU(2)_{subgr.})=Vol(U(2)/U(1))$ in the case of the group  $U(2)$,
it is necessary to introduce a parameter that allows us to
change the volume $Vol(SU(2)_{subgr.})$ without changing the volume
of the Lie algebra-identical factor  group  $Vol(U(2)/U(1))$.  In  this
$U(2)$ case, this is what is achieved  by  introducing  the  parameter
log $Vol(\bz_2)$ which allows the variation  of  the  relative  heights  of
the distribution ``peaks'' centered at the two elements of $\bz_2$.

By  introducing  an  action  giving  rise  to   extra   ``peaks''   in
$e^{S_{\Box}}$ that are centered at elements that are (by  definition)
identified in a factor group but which can contribute to the volume of
the subgroup having the same Lie algebra as the factor group, we  gain
a way of varying the volume of the subgroup {\em without}  varying  the
volume of the factor group. In the example of $U(2)$ referred to above,
the possibility of a peak in $e^{S_{\Box}}$  also  at  the  nontrivial
element of $\bz_2$ means that the volume of the subgroup  $SU(2)$  can
be made up of contributions from both peaks, whereas the volume of the
Lie algebra-identical factor group $U(2)/U(1)$ can only come from  the
peak at the identity since both elements of $\bz_2$ are identified  in
the factor group.

Relative to the approximate  $U(2)$  phase  diagram  of  Fig.  2,  the
variation of the parameter log$Vol(\bz_2)$ can be described roughly as
follows. Recall from above that in the plane defined
by  log$Vol(\bz_2)=  \log  2$,
there is only one ``peak'' in the distribution  $e^{S_{\Box}}$
(centered at the identity). In this plane, the quantity $Vol$
of  the  subgroup  is
identically twice that the factor group since,  by  definition,  there
are only half as many elements (i.e., cosets) in the factor  group  as
there are elements in the subgroup $SU(2)$ with the same Lie  algebra.
Now as the value of the  parameter  log$Vol(\bz_2)$  is  reduced,  the
pattern of fluctuations for the subgroup changes in such a way that  a
progressively larger proportion of the fluctuations  are  centered  at
the nontrivial element of $\bz_2$. These latter are not ``noticed'' by
the  factor  group  because  fluctuations  about  the   identity   and
fluctuations about the nontrivial element  of  $\bz_2$  correspond  to
fluctuations about the same coset of the factor  group  -  namely  the
identity of the factor group.  In  our  approximation,  the  parameter
log$Vol(\bz_2)$ decreases until the two peaks of  $e^{S_{\Box}}$  (one
at each element of $\bz_2$) have the same volume corresponding to half
the   volume   of   the   subgroup   $SU(2)$   (this    happens    for
log$Vol(\bz_2)=0$); i.e., the volume of $SU(2)$ is distributed equally
between the two peaks. At  just  this  point  in  the  progression  to
smaller values of log$Vol(\bz_2)$ (i.e., when log$Vol(\bz_2)=0$)), the
multiple point is encountered. This coincides with the confinement  of
the $\bz_2 \subset SU(2)$ and thereby, to the  identification  of  the
two peaks each of which contributes half the volume  of  $SU(2)$  just
prior to the confinement breakdown of  the  latter  to  $SO(3)$.  This
identification means that  the  volume  of  the  subgroup  $SU(2)$  is
reduced by a factor 2 which is just the factor by which the volume  of
this subgroup is larger than  the  Lie  algebra-identical  factor  group
$U(2)/U(1)$ in the absence  of  the  extra  parameter  log$Vol(\bz_2)$
(i.e., for the normal Manton action with  just  one  ``peak''  at  the
identity). In  particular,  with  the  help  of  the  extra  parameter
log$Vol(\bz_2)$,    the    quantities     $Vol(SU(2))_{subgr}$     and
$Vol(U(2)/U(1))$ both can have critical values for  the  same  set  of
$\beta$ parameters including the  set  at  the  multiple  point.  This
happens because, at the multiple point, the  volume  of  the  subgroup
$SU(2)$ is reduced to that of the Lie algebra-identical  factor  group
$U(2)/U(1)$. More generally, the critical  volume  associated  with  a
given Lie subalgebra is, at the multiple point,  that  of  the  factor
group obtained by identifying all elements of  discrete  subgroups  of
the center of  the  covering  (sub)group  corresponding  to  this  Lie
subalgebra.

\vspace{1cm}

In summary, this section 3.4 about the generalized action deals with
the need for more than the usual number of parameters  in
the plaquette action if one wants to make the phases corresponding  to
confinement of the various invariant subgroups  -  including  discrete
(invariant) subgroups) share a common point (i.e., the multiple point)
in the phase diagram. With our plaquette action  parameterization,  we
can show the existence of and also the  coincidence  in  one  multiple
point of phases  corresponding  to  all  invariant  subgroups  of  the
nonabelian components of the $SMG$. The invariant subgroups that we do
not consider here correspond solely to additional discrete (invariant)
subgroups of $U(1)$. The defining feature of a confinement-like  phase
for an invariant subgroup is  the  assumption  that  Bianchi  identity
constraints can be neglected for such a phase in a crude weak coupling
approximation using a mean field approximation.

At the multiple point, we are dealing with first order phase transitions;
therefore, a priori at least, our multiple point principle suffers from
lack of universality. However, the fact that a weak coupling approximation
is at least approximately applicable - even for the determination of critical
couplings - leads to the irrelevance of terms greater than second order in
Taylor expansions of the action and consequently fosters the hope
of an approximative universality.

\section{Correction due to quantum fluctuations}

In our model, the $SMG$ gauge coupling constants are to be  identified
with the couplings for the diagonal subgroup  that  results  from  the
Planck scale breakdown of $SMG^3$. While in the naive continuum limit,
the diagonal subgroup field configurations consist (by definition)  of
excitations that are identical for the $N_{gen.}=3$ copies (labelled  by
names  $``Peter",\;``Paul",\cdots$)  of  any  $SMG$  gauge  degree  of
freedom $A^b_{\mu}$, a more realistic view must take into account  that
the $N_{gen.}$ copies of $A^b_{\mu}$ in $SMG^3$: \ppm  undergo  quantum
fluctuations relative to each other. In this section  this correction is first
estimated
for a confinement-like  phase (hereby justifying a disregard of Bianchi
identities) and subsequently corrected so as to be approximately
correct for a Coulomb-like phase.

Including the effect of fluctuations of a general quantum  field  $\theta$  in
the continuum  limit  is done using  the  effective  action
$\Gamma[\theta_{cl.}]$:

\beq
\Gamma[\theta_{cl.}]=S[\theta_{cl.}]-\frac{1}{2}\mbox{Tr}(log(S^{^{\prime
\prime}}[\theta_{cl.}])). \eeq

\nin The correction to the continuum couplings that we calculate below
consists in identifying the classical continuum action $\int d^4x
\frac{-1}{4g^2}(gF_{\mu\nu}^a)^2$ with the effective
action $\Gamma$ - instead of with the lattice  action  $S$  -  in  the
naive continuum limit approximation. In calculating  this  correction,
we  ignore   nonabelian   effects   and   assume   that   the   action
$S_{_{Monte\;Carlo}}$             used             in              the
literature\cite{bachas,bhanot,drouffe} deviates only  slightly  from
the Manton action  for  which  the  Tr$log$  correction  is  simply  a
constant. The $S_{Monte\;Carlo}$ could  for  example  be  the  popular
cosine action in the $U(1)$ case.  First,  however,  we  note  that  a
change  in  the   functional   form   of   the   action   by   $\delta
S(\theta_{cl.})$
leads to a  functional  change  in
the effective action $\Gamma(\theta_{cl.})$ that differs from  $\delta
S(\theta_{cl.})$ by  a  term  proportional  to  $\mbox{Tr}\frac{\delta
S^{^{\prime                        \prime}}[\theta_{cl.}]}{S^{^{\prime
\prime}}[\theta_{cl.}]}$:

\beq  \delta   \Gamma[\theta_{cl.}]=\delta   S[\theta_{cl.}]+\frac{1}{2}
\mbox{Tr}(\delta(\log (S^{^{\prime  \prime}}[\theta_{cl.}])))=  \delta
S[\theta_{cl.}]+\frac{1}{2}         \mbox{Tr}(\frac{\delta(S^{^{\prime
\prime}}[\theta_{cl.}])}{S^{^{\prime
\prime}}[\theta_{cl.}]}) \eeq

\nin  But  as   we   are   assuming   that   the   variation   $\delta
S[\theta_{cl.}]$ is done relative to the Manton action, we have

\beq \frac{1}{2}         \mbox{Tr}(\frac{\delta(S^{^{\prime
\prime}}[\theta_{cl.}])}{S^{^{\prime
\prime}}_{Manton}[\theta_{cl.}]}) = \frac{1}{2}         \mbox{Tr}(S^{^{\prime
\prime}}[\theta_{cl.}]\langle
(\theta-\theta_{cl.})^2\rangle) \eeq

\nin     where      we      have      used      that      $S^{^{\prime
\prime}}_{Manton}[\theta_{cl.}]    \propto     \langle
(\theta-\theta_{cl.})^2\rangle^{-1}=\;const.$     and     that,      up      to
a  constant,  $\delta(S^{^{\prime  \prime}}[\theta_{cl.}])=S^{^{\prime
\prime}}[\theta_{cl.}]$ (modulo a constant).

Neglecting
nonabelian effects, we generalize  this  result  to  nonabelian  gauge
groups and write it more concretely using $U=e^{i\theta^at^a}$ and
$S[U]=\sum_{\Box}S_{\Box}(U(\Box))$:

\begin{equation}   \Gamma    [U_{cl.}]=S[U_{cl.}]    +
\frac{1}{2}\Delta S_{\Box}(U_{cl.}(\Box))\langle (\theta_{\Box}^a(\Box)-\theta
_{\Box\;cl.}^a(\Box))^2\rangle
\;\;  (\mbox{summation
over}\;a) \label{gamma}\end{equation}





\nin We have for the Laplace-Beltrami operator

\begin{small}

\beq \frac{1}{2}\Delta    S(U(\Box))    \stackrel{def.}{=}
\lim_{\epsilon \rightarrow 0^+}\frac{\int  d^{N^2-1}\!f\;\;
\exp(-\frac{1}{2\epsilon}  f_a^2)(S(U  \cdot   e^{if_bt_{b}})-S(U))}{\int
d^{N^2-1}\!f\;\; \exp(-\frac{1}{2\epsilon} f_d^2)f_e^2} \;\; (\mbox{sum  over
}a,b,d,e). \eeq

\end{small}

\nin where $f_a$ and $t_{a}$  denote  respectively
the $a$th Lie algebra component and Lie algebra generator.
Upon expanding (in the representation $r$) the exponential
$\exp(if_b T_{b,\,r})$ representing $\exp(if_b t_{b})$
the argument  of  which  is  assumed  to  be  small  inasmuch  as  the
$f_b$ are assumed to be small,
there obtains

\begin{small}

\beq =  \lim_{\epsilon  \rightarrow  0^+} \sum_r \frac{\beta_r}{d_r}
\frac{\int   d^{N^2-1}\!f\;\;
\exp(-\frac{1}{2\epsilon} f_a^2)\mbox{Tr}(U \cdot (-\frac{1}{2}f_bf_cT_{b,\,r}
T_{c,\,r}))}{\int      d^{N^2-1}\!f\;\;      \exp(-\frac{1}{2\epsilon}
f_d^2)f_e^2} \;\;{\footnotesize (\mbox{sum  over  }a,b,c,d,e)}\eeq


\beq =\sum_r\frac{\frac{\beta_r}{d_r}Tr_r(U\cdot
(-\frac{1}{2}(T_{b,\,r})^2))}{N^2-1} \;\;(\mbox{sum over }b) \eeq

\end{small}

\nin where we have expanded the plaquette action  in  characters:  for
the representation  $r$  of  dimension  $d_r$  the  character  $\chi_r$
is given by $\chi_r=\mbox{Tr}_r(U(\Box))$ and

\beq S(U(\Box))=\sum_r \frac{\beta_r}{d_r}\mbox{Tr}_r(U(\Box)).     \eeq

\nin We have

\beq                        \frac{1}{2}                         \Delta
S(U(\Box))=\sum_r-\frac{1}{2}\frac{\beta_r}{d_r}
Tr_r(U(\Box))\frac{C^{(2)}_r}{N^2-1} \eeq


\nin where $C^{(2)}_r$ is the quadratic Casimir for the representation
$r$.         The         Casimir         is         defined         as
$C^{(2)}_r\bunit_r\stackrel{def.}{=}\sum_b(T_{b,\;r})^2    $.
For the groups  $SU(2)$  and  $SU(3)$,
the Lie algebra bases in the  fundamental  (defining)  representations
are      taken      respectively       as       $T_{b,\;r=f}=T_{b,\;r=
\underline{\sbtwo}}=\frac{\sigma^b}{2}$     and     $T_{b,\;     r=f}=
T_{b,\;r=\underline{\sbthree}}=\frac{\lambda^b}{2}$. The subscript $f$
denotes the fundamental representation; $\sigma^b$ and $\lambda^b$ are
the   Pauli   and   Gell-Mann   matrices   with   the    normalization
$\mbox{Tr}(\frac{\sigma^a}{2}\frac{\sigma^b}{2})=\frac{\delta_a^b}{2}$
and                  $                   \mbox{Tr}(\frac{\lambda^a}{2}
\frac{\lambda^b}{2})=\frac{\delta_a^b}{2}$. With this basis convention,
and with the left-handed  quark  doublet  field  as  an  example,  the
covariant derivative is

\beq
D_{\mu i\;\;\;\alpha}^{\;\;\;\;j\;\;\,\beta}=
\partial_{\mu}\delta_i^{\;\;j} \delta_{\alpha}^{\;\;\beta}
-ig_2A_{\mu}^b\frac{(\sigma^b)_{ i}^{\;\;j}}{2}\delta_{i}^{\;\;j}
-ig_3A_{\mu}^b\frac{(\lambda^b)^{\;\;
\beta}_{\alpha}}{2}\delta_{\alpha}^{\;\;\beta}
-ig_1\frac{1}{6}A_{\mu}\delta_i^{\;\;j} \delta_{\alpha}^{\;\;\beta}.
\eeq

\nin where the index $b$ labels Lie algebra  components,  the  indices
$i,\;j$  label  matrix  elements  of   the   ($2$-dimensional)  fundamental
representation of $SU(2)$, and the indices $\alpha,\;\beta$ label  the
matrix elements of the ($3$-dimensional) fundamental representation of
$SU(3)$. The factor $\frac{1}{6}$ in  the  last  term  is  the  $U(1)$
quantum  number  $\frac{y}{2}$  where  $y$  is  weak  hypercharge;  the
convention used is $Q=\frac{y}{2}+I_{W_{3}}$.

The above convention for the generators of $SU(2)$ and $SU(3)$ in  the
fundamental    representation    $f$    leads     to     a     Casimir
$C^{(2)}_f=\frac{N^2-1}{2N}$  for  an  $SU(N)$  group.  From  this  it
follows that, for the adjoint representation (denoted by $adj.$),  the
Casimir $C^{(2)}_{adj.}$ for an $SU(N)$ group is given by
$C^{(2)}_{adj.}=N$.





%





Ignoring Bianchi identities, we get for the deviations

\beq
\langle( \theta_{Peter}- \theta_{diag.})^2_a          \rangle=
\frac{N^2-1}{2}                                                          (
\frac{1}{2} \sum_r  \beta_r \frac{C^{(2)}_r}{N^2-1})^{-1}
(\mbox{sum over }a) \mbox{   (confinement phase)}\eeq







\[=-\{\frac{\beta_f}{d_f} Tr_f(U_{diag.}(\Box))C^{(2)}_f+
\frac{\beta_{adj.}}{d_{adj.}}
Tr_{adj.}(U_{diag.}(\Box))C^{(2)}_{adj.}\}\frac{1}{2}
\frac{N^2-1}{\beta_fC^{(2)}_f+ \beta_{adj.}C^{(2)}_{adj.}} \]


Letting the sum over representations run  only  over  the  fundamental
(=defining) and adjoint representations labelled  respectively  by  the
subscripts $f$ and $adj.$ (the only representations used in the Monte Carlo
runs of references \cite{bachas,bhanot,drouffe}),  we  get  for  the
effective action (\ref{gamma})

\begin{equation} \Gamma (U_{diag.}(\Box))= \end{equation}
\[ = \frac{\beta_f}{d_f} Tr_f(U_{diag.}(\Box))
(1-\frac{C^{(2)}_f(N^2-1)}{2( \beta_fC^{(2)}_f+ \beta_{adj.}C^{(2)}_{adj.})})+
 \frac{\beta_{adj.}}{d_{adj.}} Tr_{adj.}(U_{diag.}(\Box))
(1- \frac{C^{(2)}_{adj.}(N^2-1)}{2(\beta_fC^{(2)}_f+
\beta_{adj.}C^{(2)}_{adj.})}) \]


\nin So with the continuum correction we have  to make the replacement

\beq \beta_r \rightarrow \beta_r(1-\frac{C^{(2)}_r(N^2-1)}
{2\sum_{\hat{r}} \beta_{\hat{r}}C^{(2)}_{\hat{r}}}) \;\;
\mbox{(for ``confinement phase'')} \label{eqn31}\eeq


This expression for the effective action has been obtained  using  the
approximation  that  all  plaquette  variables  can  be  regarded   as
independent (i.e., Bianchi identities have been disregarded).  This
approximation is appropriate for the confinement  phase.  However,  as we
are interested in criticality as approached  from  the  Coulomb  phase
(i.e., Coulomb phase in our scale dependent sense),   we want
the quantum fluctuation  correction  in  this  phase
where Bianchi identities must be respected.  These  identities  reduce
the number of degrees  of  freedom  per  plaquette  that  can
fluctuate independently by a factor  2.  In going to  the  Coulomb  phase,  the
continuum-corrected $\beta_r$ is modified  as
follows:

\beq \beta_r(1-\frac{C^{(2)}_r(N^2-1)}
{2\sum_{\hat{r}} \beta_{\hat{r}}C^{(2)}_{\hat{r}}})_{confinement}\rightarrow
\beta_r(1-\frac{C^{(2)}_r(N^2-1)}
{4\sum_{\hat{r}} \beta_{\hat{r}}C^{(2)}_{\hat{r}}})_{Coul.\;phase}\label{eqn32}
\eeq

\nin Without the continuum correction, we have for the fine  structure
constants at the multiple (i.e., triple ) point

\beq \frac{1}{ \alpha_{triple\;point,\,no\,cont.}}
=4\pi\sum_r \frac{C^{(2)}_r \beta_{r,\,triple\;point}}{N^2-1}
\;\;\mbox{(naive continuum limit)}. \eeq

\nin With the continuum-corrected $\beta_r$ in the  Coulomb  phase  we
have for the fine structure  constants  at  the  triple point

\begin{small}

\beq\frac{1}{\alpha_{_{triple\;point, \,cont.}}}=
4\pi\sum_r \frac{C^{(2)}_r}{N^2-1} \beta_{_{r\,triple\;point}}
(1-\frac{C^{(2)}_r(N^2-1)}
{4\sum_{\hat{r}}\beta_{_{ \hat{r},\,triple\;point}}C^{(2)}_{\hat{r}}})= \eeq
\[ 4\pi\sum_r\frac{C^{(2)}}{N^2-1} \beta_{r\;triple\;point}(1-\pi
C^{(2)}_r \alpha_{_{triple\;point,\;no\;cont.}}) \label{corbian} \]

\end{small}



\section{Calculation of critical couplings at Planck scale}

It can be argued that at the multiple point of the phase  diagram  for
the whole $SMG^3=SMG\times SMG\times SMG$, the nonabelian  (plaquette)
action parameters for each of the three cartesian product factors take
the same values as at the multiple point  for  a  single  gauge
group $SMG$. This allows us to determine  the  multiple  point  action
parameters for the  gauge  group  $SMG^3$  from  a  knowledge  of  the
multiple point action parameters for just one of the $SMG$ factors  of
$SMG^3$.
Accordingly, we can calculate the multiple point critical couplings
from the couplings for the isolated $SU(2)$ and $SU(3)$ groups. To this end,
we have
used  figures  from  the
literature  \cite{bachas,bhanot,drouffe} to graphically extract the
coordinates $(\beta_f,\beta_{adj.})_{triple\;point}$ of the
triple point:

For $SU(2)$:  $(\beta_f,\beta_{adj.})_{triple\;point}= (0.54, 2.4)$

For $SU(3)$:  $(\beta_f,\beta_{adj.})_{triple\;point}= (0.8, 5.4)$

\nin  The  calculation  of  $\alpha_2^{-1}$  and  $\alpha_3^{-1}$  are
presented in Tables  3  and  4  respectively.  In  these  tables,  the
subscripts ${adj.}$  and  $f$  denote  respectively  the  adjoint  and
fundamental representations of the groups considered.

In order to get an idea of the order of magnitude of the error involved
in estimating the average over the Laplace-Beltrami of the plaquette action
only to next to lowest order, we make use of the fact that we can calculate
this average to all orders in the case of a $\cos\theta$ action for a $U(1)$
gauge theory. In this case the
averaging is  readily  performed  and leads  to  an
exponential for which the first terms of a Taylor  expansion  coincide
with the terms we  calculated using \ref{corbian}.
This  suggests  that  also  in  the
nonabelian cases it might be quite reasonable to ``exponentiate''  our
``continuum'' corrections and subsequently use  the  change  made  by
such a procedure as a crude estimate of the error due to our omission
of the second order perturbative terms. By exponentiated continuum corrections
we mean by definition that, instead of the replacements \ref{eqn31} and
\ref{eqn32}, we use respectively

\beq \beta_r\rightarrow \beta_r\exp(-\frac{C^{(2)}_r(N^2-1)}
{2\sum_{\hat{r}} \beta_{\hat{r}}C^{(2)}_{\hat{r}}}\;\;\;\mbox{(for
``confinement'')}\eeq

\nin and

\beq \beta_r\rightarrow \beta_r\exp(-\frac{C^{(2)}_r(N^2-1)}
{4\sum_{\hat{r}} \beta_{\hat{r}}C^{(2)}_{\hat{r}}} \;\;\;\mbox{(for ``Coulomb''
phase)} \eeq

\nin   As evidenced by Tables  1  and
2, this exponentiation yields a change of the  order  of  one  unit  in
$1/\alpha_{crit.,\;cont.}$ $\approx$ 20 from which we can estimate the
uncertainty due the neglect of higher order  terms  as  being  of  the
order of say 5 \%. While we are on the subject of uncertainty, we should
mention that we have even ignored a  relatively  little  term\cite{cs}
having the same order  in  perturbation  in  $1/\beta$  as  the  above
continuum correction. In the abelian case, this term  is  1/4  of  the
continuum correction. Because this term vanishes for  $SU(N)$  in  the
large $N$ limit, it is even smaller for the couplings for $SU(2)$  and
$SU(3)$.

Since our deviations from the experimental couplings  extrapolated  to
the Planck scale\cite{kim,amaldi}  are of the same order of magnitude as
the uncertainty
in the Monte Carlo data and the
uncertainty due to chopping off the higher order continuum corrections,
a calculation of the next  order  corrections
and increased accuracy in the calculations
are called for in order to determine if our deviations are significant.

\section{Proposed model for  the  stability  of  the  multiple
point}


We propose a  mechanism
for the stability of the multiple point that is based on a  model
which could be called a ``nonlocal gauge glass model''\footnote{The term
``gauge
glass'' was appropriately coined by Jeff Greensite  by  analogy  to  a
spin glass which is so named because the ``frozen  in''  structure  is
reminiscent of that of glass.}.
which is very  much  inspired  by
the project of ``random dynamics\cite{randyn}''.
The
essential feature is the influence of a bias effect that can occur  in
the presence of a plaquette (or multi-plaquette) action the functional form
of which is taken to be quenched random.
This could  mean  that  for  each
Wilson  loop $\Gamma$,  the  coefficients  (called  the  ``$\beta$'s'')  in say
 a
character expansion of the Wilson loop action are fixed at the  outset
as random values and remain fixed during the evaluation of the functional
integral. While translational invariance is broken at least at  small
scales because a different set of random $\beta$'s  is  associated  with
each Wilson  loop,  it  is  presumably  regained  at least
approximately in  going  to  large
distances inasmuch as it is assumed that the statistical  {\em distribution}
of quenched random variables is translationally invariant.


Randomly weighted terms in the  action  from  the  different  Wilson
loops would on the average contribute nothing to the  inverse  squared
coupling were it not for the {\em bias}:
the vacuum dominant value of a  Wilson
loop variable (a point in the gauge  group)  is  correlated  with  the
values of the quenched random coefficients for the Wilson  loop  under
consideration. This correlation comes about because the
vacuum field  configuration\footnote{Note that we envision
a relatively complicated vacuum state in which the link or rather plaquette
variables fluctuate around other elements than the unit element. However,
these ``other elements'' must necessarily be elements of the center
if ``collapse'' ($\approx$ Higgs-like behavior) is to be avoided; this may
require a connected center\cite{cen} for the group that extends almost
densely over the group.} adjustments  resulting
from the tendency to approximately  maximize  the  exponential  of  the
action $\exp(S)$ as a function of link variables  will  concurrently  tend  to
make the second derivative  w. r.  t.
Wilson loop variables of $\exp(S)$ more negative.


In the simplest model, the gauge glass we use is rather strongly  {\em
nonlocal} because we assume that the quenched random contributions  to
the  action  are  not  restricted  to  contributions  from   elementary
plaquettes, but in principle include {\em all} Wilson loops.  If  this
should lead to problems with locality,  we  can  postulate  that  only
loops up to some finite size are present in the action since the
most crucial prerequisite for the bias mechanism is the  inclusion  of
many Wilson loops with the size distribution being of  only  secondary
importance.

The bias effect can be formulated as
an additional term in the Callan-Symanzik $\beta$-functions (in
addition to the normal renormalization group contribution). To see this,
envision a series of calculations
of the effective  couplings $g(\mu)$ for successively larger
inverse energies $\mu^{-1}$.
For each value of $\mu^{-1}$, Wilson loops of size up $\mu^{-1}$ are
included in computing $g(\mu)$; therefore a calculation of $g(\mu)$
includes more and more Wilson loops in going towards the infrared.
The inclusion of progressively longer
and longer  loops  takes  place  in  a  background  field  made  up  of
contributions  from  the  already  included  smaller  loops  that  are
approximately described as a  background  continuum  Lagrange  density
$-\frac{1}{4g^2(\mu)}F_{\mu\nu}^2$. This process, in which the  coupling
$g(\mu)$ becomes smaller and smaller the more loops it  accounts  for,
culminates in $g(\mu)$ attaining  the  critical  value  whereupon  the
influence  of  additional  loops  on  the  vacuum   configuration   is
drastically  diminished  because  the   transition   to   a   $g(\mu)$
corresponding to the Coulomb phase leads  to  a  vacuum  configuration
that is much less readily influenced than in  the  confinement  phase.
Contributions from larger Wilson loops are no longer  correlated  with
the vacuum dominant field configuration that is almost solely  determined
by  the  Wilson  loops  of  smaller  spatial   extent.   Without   the
``protection'' of the bias effect, the contributions from these larger
loops cancel out on the average because of the assumed  randomness  in
the signs of action terms with the result that the effective couplings
will  no longer be modified much
by the inclusion of  larger  Wilson  loops  that
show up in going to larger length scales.


The  variation  of  the
effective coupling due to the bias effect might formally  be  included
in a generalized Callan-Symanzik $\beta$-function. (actually we mean a
multicomponent  vector  of  generalized  $\beta$-functions  with   one
component for each  parameter  of  a  single  plaquette  action  of  a
course-grained  lattice  at  the  scale  $\mu$).   These   generalized
$\beta$-functions (i.e., the components of the vector  of  generalized
$\beta$-functions) contain contributions that take into  account  that
the part of the Lagrangian of the theory that is used to define  gauge
couplings $g(\mu)$ is changing as we go to larger length scales.  That
this change has a nonvanishing average effect on the couplings is  due
to   the   bias   effect.   These   extra   contributions    to    the
$\beta$-functions, which are in addition to the normal renormalization
group effects, make the generalized $\beta$-functions explicitly scale
dependent.  Specifically,  we  envision  rapid   variations   of   the
$\beta$-functions as the bias effect is drastically  weakened  at  the
transition to a Coulomb-like phase. If  the  $\beta$-functions  become
zero, this would result in an infrared attractive fixed point near the
phase transitions at the multiple point.


An important point is that  multiple point
criticality is implied by  almost  any
mechanism that drives a gauge coupling to a critical value  because  a
mechanism that seeks out the critical coupling for  some  gauge  group
will probably function in the same way for all invariant subgroups  of
a gauge group. But this is tantamount to seeking out the multiple
point which by definition is the point or surface in the phase diagram
at the borderline between confining  and  Coulomb  phases  for  all
invariant subgroups. In particular, our model as outlined above would,
imply that  Wilson  loop  contributions  $\prod_{\;\link  \;  \in
\Gamma}U(\link)$ depending only on the cosets  in  $G/H$  w.r.t.  some
invariant subgroup $H$ would become very ineffective  in bringing about  a
further increase in  the  inverse  squared  couplings  (for  the  \dof
corresponding to the factor group)  once  it  is  only  the  invariant
subgroup $H$ of the group $G$ that  remains  ``confining'';  in  other
words,  the  couplings  for  $G/H$  stop  falling  (in   the   crudest
approximation) once $G/H$ ``reaches the Coulomb phase''.

Several alternatives to the nonlocal gauge glass explanation  for  the
stability of  the  multiple  point  (assuming  it  exists)  have  been
considered. We have for example in previous work, prior to the  advent
of the principle of multiple point criticality, used  the  entropy  as
the quantity to be maximized in the  predictions  of  gauge  couplings
from criticality. In this earlier work, we have for the nonabelian Lie
subgroups of the $SMG^3$ considered criticality only  w.r.t.  the  Lie
subgroups $SU(2)$ and $SU(3)$ and not for criticality  w.  r.  t.  the
$\bz_2$ and $\bz_3$ discrete  invariant  subgroups  of,  respectively,
$SU(2)$ and $SU(3)$. We found that the entropy, calculated  to  lowest
order, was to first approximation constant on an interface  of  finite
extent that separated the totally Coulomb-like and totally confinement
phases in the parameter space of the $SMG$ phase diagram.  In  effect,
this interface prohibited other  \pcps  from  meeting  at  a  multiple
point. We now find that the addition of action  parameters  that  also
allow the discrete invariant subgroups to become critical  results  in
the shrinking of this interface into a point that coincides  with  the
multiple point. This can be expected to affect the entropy because, at
the multiple point, we are also on the verge of  confinement  for  the
groups $\bz_2$ and $\bz_3$. This means that  a  small  change  in  the
appropriate action parameters can bring  about  a  transition  from  a
Coulomb-like phase to a  confinementlike  phase  with  the  difference
between  the  two  being,  for  example,  defined  by  the  respective
perimeter and area law decay of Wilson loops (for charges 1/2  or  1/3
in the case of $\bz_2$ and $\bz_3$ respectively).

In the action  parameter  space  that  includes  parameters  that  can
be adjusted so as to have criticality w. r. t. the discrete  invariant
subgroups $\bz_2$ and $\bz_3$, the entropy is constant to lowest order
along a (hyper)surface separating the  totally  Coulomb  from  totally
confined phases. But a calculation to next order appears  to  lead  to
the conclusion that the entropy is not maximum  at  the  multiple
point thereby obviating the idea of maximum entropy as an  explanation
for the multiple point.

However, it can be claimed that  the  multiple point  is  such  a
characteristic ``corner'' of the phase diagram that  it  is  extremely
likely that there is some relevant physical quantity  or  property  that
is extremized at this point. A possible scenario that  might  in  part
rescue the maximum entropy idea is that, at the  multiple point,
there are strong fluctuations along the discrete  subgroup  directions
of the gauge group that, for given entropy, might be very  effective  in
preventing potential Higg's fields from  bringing  the  model  into  a
Higg's phase.  In  other  words,  the  entropy  that  comes  from  the
``discretized'' lack of knowledge (as to which element of the discrete
invariant  subgroups  in  the  neighborhood  of  which  the  plaquette
variable takes  a  value)  may  function  better  in  suppressing  the
tendency for ``Higgsning'' than the same  amount  of  entropy  arising
from  fluctuations  within  the  individual   Gaussian   distributions
$e^{S_{Manton}}$ centered at the elements of $\bz_2$  and  $\bz_3$.  If
this were true, one might use Higgs suppression as the property  to
be optimized at the multiple point.

Yet another admittedly rather speculative approach to  explaining  the
multiple point suggest  that  the  functional  integral  for  the
partition function in baby universe theory should have a maximal value
at the multiple point\cite{baby}.

Meaningful continuum couplings for lattice gauge theories do not exist
for   couplings   that   exceed   the    critical
values\cite{ngen,sur}. This is  corroborated  by  the  observation\cite{sur}
that Mitrushkin\cite{mit} only formally  obtains  a  strong  continuum
coupling in the Coulomb phase.

In summary, we have  in  this  section  supplemented  the
postulate  of  the  principle  of  multiple  point  criticality   with
proposals as to how a stable Planck  scale  multiple  point  might  be
realized. To this end, we described a gauge glass model inspired
by random dynamics. This model, which uses a quenched random action, has a
bias causing weaker couplings that is discontinuously
diminished at the multiple point. This leads to a zero of a generalized
Callan-Zymanzik $\beta$ function  thereby establishing the multiple point
as an approximate ``infrared stable'' fixed point.

The speculative nature of these arguments in no way detracts
from the most important justification for the principle which  is  the
noteworthy phenomenological success.


\section{Is our ``anti-unified multiple point model'' consistent with
established results?}

Since the breakdown of our ``anti-unified'' $SMG^3$ to the diagonal
subgroup (that is identified with the standard model group) occurs  at
a scale close to the Planck scale,  we  predict  pure  standard  model
results with an accuracy that is even greater than $SU(5)$ G.U.T. with
or without supersymmetry  (both  of  which  have  (on  a  log  scale)
a slightly lower unification scale than the Planck scale).

Since our good agreement  for  coupling  constant  values  would  not
tolerate  the  survival  of  supersymmetry  down   to   experimentally
accessible  scales,  we  accordingly   predict   that   supersymmetric
partners, if they exist at all,  cannot,  in  practice  at  least,  be
observed. What could be tolerated in our model, which a priori assumes
a desert in the renormalization  group  extrapolation  to  the  Planck
scale, are fields that would not appreciably affect the renormalization
group development of the finestructure constants.

As alluded to in the Introduction,  our  model  also  predicts,  as  a
consequence of the  multiple  point  criticality,  light  (compared  to
Planck scale) scalars provided that this principle is extended to mean
that Nature seeks parameter values that also are on the boundaries  of
Higgs phases of  various  types.  Assuming  weak  first  order  (i.e.,
approximately second order) phase transitions at these boundaries, the
predicted masses would presumably in most cases be a  few  orders  of
magnitude under the Planck mass. If  some  of  the  phase  transitions
happen to be very nearly second order, the corresponding scalars might
turn  out  to  have  experimentally   accessible  masses.   For   the
$\beta$-function  as  it  pertains  to  the   gauge   couplings,   the
contributions from a scalar is  smaller  than  from  a  fermion  or  a
Yang-Mills particle  in  the  same  representation.  Therefore,  these
scalars are not expected  to  influence  our  predictions  appreciably
provided only few of them have  very  small  masses  compared  to  the
Planck scale.

Proton decay at the $SU(5)$ unification scale is predicted to be absent in
our model because we simply do not have $SU(5)$ in our model.
Above the $SMG^3$ breakdown scale (still approximately the Planck  scale)
where also in our model almost anything can happen,  baryon  violation
may take place. The fact that baryon violation would  first  occur  at
the Planck scale would  imply  a  long  proton  lifetime  compared  to
present experimental limits if in fact it decays at all.


With the exception of predictions of parameters of the standard model,
our model has only  very  limited  possibilities  for  conflicts  with
present day experimental results  because  present  experiments  agree
with the same pure standard model (including  the  desert!)  which  we
essentially predict almost up to the Planck scale. However,  from  the
viewpoint of astrophysics, our model may make some predictions:  since
the homotopic group

\[ \Pi_1(SMG^3/SMG_{diag})\;\overline{\simeq}\;  \Pi_1(U(1)^3/U(1)_{diag})
\;\overline{\simeq}\; \Pi_1(U(1)^2)\; \overline{\simeq}\; \bz^2 \]

\nin of the space of Higgs vacuum states could lead to three types  of
stable cosmic strings that can branch into each other, a scenario with
galaxy formation caused by cosmic  strings  would  have  some  details
modified in a welcomed way\cite{bjones}.

Although still suppressed compared to a Planck scale  energy  density,
the cosmic strings of our model  are  expected  to  have  high  energy
density compared to strings related to an $SU(5)$-scale.

New physical predictions may follow from our more
general plaquette actions $S_{\Box}$  having  parameters  that  can  be
adjusted so as to have distributions $e^{S_{\Box}}$ of  varying  width
that are concentrated  within  say  an  invariant  subgroup
$\bz_N$ of the $SMG$ (independent  of  the  distribution
widths  along   other   invariant
subgroups). Roughly our multiple point principle means that both
discrete and Lie subgroups are simultaneously on  the  border  between
confinement and Coulomb. Now discrete gauge theories are really string
theories in the sense that - in the Coulomb  phase  -  magnetic  field
lines are (dual) strings. In the confinement phase, the magnetic field
lines condense in the  vacuum  and  electric  field  lines  appear  as
strings. Even for first order phase transitions, the string tension of
the magnetic/electric flux lines is at a minimum infinitesimally close
to the transition (but still in Coulomb/confinement phase) because  it
approaches zero in the opposite phase  (i.e,  the  confinement/Coulomb
phase). If the discrete group deconfinement transition  should  happen
to be second order or very weak first order (e.g., this might  be  the
case for a $\bz_N$ with sufficiently large $N$), the  string  tensions
at the transition would be zero (or  small).  Thinking  of  a  $\bz_N$
discrete subgroup, magnetic flux concentrated  in  flux  tubes  having
units of $2\pi/N$ could have very low energy. These flux  tubes  would
appear as strings between monopoles of charge $2\pi/N$ or form  closed
loops  that  could  conceivably  be  found  at  energies  aspired   to
experimentally.

A speculative connection between our model with the gauge group $SMG^3$ and
experimentally observable data can be established by adding the assumption
that the mass hierarchy of quarks and leptons is due to an approximate
conservation of the quantum numbers of our gauge model. This amounts to having
small expectation values for the Higgs fields used to break
$SMG^3$ to $SMG$.
A recently performed\cite{gap} analysis of this type lead to the conclusion
that the high top quark mass poses a problem for our model unless the model
is endowed with an additional conserved quantum number (for an extra $U(1)$).

\section{Conclusion}

In this paper, we predict the  values  of  the  fine  structure
constants realized by Nature at the Planck scale  for  the  nonabelian
gauge groups $SU(2)$ and $SU(3)$ that agree with experiment to  within
our calculational accuracy. The prediction is that  these  coincide  with  the
continuum limit of the diagonal subgroup  couplings  corresponding  to
multiple point values of the action  parameters  of  a  lattice  gauge
theory with  the  nonsimple  gauge  group  $SMG^{N_{gen.}}=SMG^3=  SMG
\times SMG \times SMG$. Here $N_{gen.}=3$ is the number of  quark  and
lepton generations and $SMG$ denotes the standard model group.  It  is
assumed in our model that the more  fundamental  gauge  group  $SMG^3$
breaks down at the Planck scale to the diagonal subgroup which  is  by
definition identical with the standard model group. The multiple point
in the phase diagram of a lattice  gauge  theory  having  a  nonsimple
gauge group is a point at  which  there  are  critical  values  for  a
maximum number of the action parameters; at this point,  infinitesimal
variations of these parameters can provoke ``confinement-like phases''
corresponding to each (or at least many) of the invariant subgroups of
the gauge group. The criticality referred to here is that for the bulk
phase transition  of  a  (Euclideanized)  lattice  gauge  theory.  The
experimental values for the W-couplings and the  QCD  gluon  couplings
after extrapolation to the Planck scale\cite{kim,amaldi} are  compared
with multiple point couplings obtained by correcting data  from  Monte
Carlo results \cite{bachas,bhanot,drouffe}. When these corrections  to
the critical couplings at the multiple point are taken  into  account,
we find that, to within a 7\% accuracy,  there  is  a  desired  factor
three  deviation  between  the  values  for   extrapolated   continuum
couplings for $SU(2)$ and $SU(3)$ and the triple point  couplings  for
single $SU(2)$ and $SU(3)$ gauge groups. It is remarkable that this is
true with an accuracy that is comparable to the uncertainty that comes
about due to our having only performed the corrections connecting  the
lattice action $\beta$-parameters with the continuum couplings to next
to lowest order in perturbation theory. That is to say that, to lowest
order perturbation in  $\beta^{-1}$  for  continuum  corrections,  the
exact validity of the factor 3 between the  extrapolated  experimental
coupling values and the multiple point values cannot be  excluded.  As
regards the factor 3, it should also be mentioned that our accuracy of
$7\%$ is on the  verge  of  being  good  enough  to  be  taken  as  an
indication  that  the  ratio  $\alpha_{multi.\;point}/\alpha$  has  an
integer  value.  Having  this  ``3''  amounts  to  saying   that   the
extrapolated fine structure constants are just three times as small as
the couplings corresponding to the  multiple  point  action  parameter
values of the lattice gauge theory with a single $SMG$. But the factor
3 is ``explained'' by our postulate of the Planck scale  breakdown  of
the ``anti-unified'' gauge group $SMG^3$  to  the  diagonal  subgroup.
This  follows  because  the  diagonal  subgroup  has  inverse  squared
couplings that are the sum of the inverse squared couplings  for  each
of the $N_{gen}=3$ $SMG$ factors in $SMG^3$. A possible mechanism  for
this breakdown is that  of  ``confusion''\cite{confusion,gaeta}  which
causes the gauge group $SMG^3$ to break down to the diagonal  subgroup
$SMG_{diag}$; the latter consists of those group elements  of  $SMG^3$
that are left invariant under the automorphism group (the latter being
the group of permutations  of  the  $SMG$  factors  in  the  cartesian
product $SMG^3$).



It can be argued that at the multiple point of the phase  diagram  for
the whole $SMG^3=SMG\times SMG\times SMG$, the nonabelian  (plaquette)
action parameters for each of the three cartesian product factors take
the same values as at the multiple point  point  for  a  single  gauge
group $SMG$. This allows us to determine  the  multiple  point  action
parameters for the  gauge  group  $SMG^3$  from  a  knowledge  of  the
multiple point action parameters for just one of the $SMG$ factors  of
$SMG^3$. We mention in passing that this is not  so  for  the  abelian
couplings; here it is necessary to seek  the  value  at  the  multiple
point for the whole $U(1)^3$. The reason for this is  related  to  the
possibility  that,  for  $U(1)$,  there   can   be   ``mixed''   terms
$F_{\mu\;\nu}^{Peter}  F^{\mu\;\nu\;Paul}$  even  in   the   continuum
lagrangian as opposed to the  case  for  nonabelian  \dof  where  only
quadratic terms $F_{\mu\;\nu}^{Peter}F^{\mu\;\nu\;Peter}$ appear  (the
names ``$Peter$'' and ``$Paul$'' label a pair of the $SMG$ factors  of
$SMG^3=SMG_{Peter}\times    SMG_{Paul}\times    SMG_{Maria}$).    This
complication may well lead to a phenomenologically  desired  factor  6
for the abelian coupling instead of the  factor  3  obtained  for  the
nonabelian couplings.


We  think  that  our  accuracy  of  7\%  or  so  for  the   (inverse)
fine structure constants at the Planck scale is already so  impressive
that it strongly suggests that there may be some truth behind the part
of our model that is relevant for obtaining this result. Furthermore,
the  validity
of our arguments is presumably rather insensitive to  whether  we  use
precisely the gauge group $SMG^3$; indeed it is
presumably sufficient that the nonabelian groups $SU(2)$ and
$SU(3)$ are imbedded
as diagonal subgroups in whatever the unifying or anti-unifying  group
might be. As long  as  the  nonabelian  subgroups  appear  as  the
diagonal subgroup of three isomorphic subgroups, it would not affect
our result if there were other cartesian product  factors  involving
again the same or other subgroups in the  anti-unifying  group.
However, if the gauge group were embedded in a larger simple  group  -
such     as     Georgi-Glashow\footnote{It     is      an      amusing
coincidence\cite{korfu} however that the value  of  the  SUSY  $SU(5)$
coupling at the unifying scale say is not so far  from  being  at  the
multiple point for a  single  $SU(5)$.}$SU(5)$,  it  would  impair  the
agreement with our model as would supersymmetry if it were not  broken
close to the Planck scale so as not  to  invalidate  our  use  of  the
``desert'' extrapolation  of  experimental  couplings  to  the  Planck
scale.

We offer a speculative theoretical explanation for  the  stability  of
the multiple point the essence of which is that the inclusion of  more
and more Wilson loops with quenched random coefficients in the  action
makes the continuum coupling decrease rapidly as long as the theory is
in the ``confinement-like'' phase. However, at  the  phase  transition
(i.e., multiple point), we speculatively predict  that  this  tendency
for couplings to continue to become weaker is  so  rapidly  attenuated
that it suddenly can be compensated by  normal  renormalization  group
effects already at energies of the order of  the  Planck  scale.  This could
lead to a zero of the effective $\beta$-function (actually one should
think of a $\beta$-function vector with one component for each of  the
action parameters) for  coupling values close to those at the  multiple
point which then effectively functions as an ``infrared stable'' fixed
point (really an ``infrared'' attractive surface).

While we purport to have demonstrated the  phenomenological  relevance
of the principle of multiple  point  criticality  for  the  nonabelian
couplings, we point out that there are as yet unresolved problems with
the group $U(1)$ because of the infinite number of invariant subgroups
and corresponding phases expected for  $U(1)^3$.
However, we have made recent progress suggesting  that,  while  it  is
presumably not possible to find a multiple point corresponding  to  an
action parameter set with critical values for all invariant  subgroups
of $U(1)^3$, a candidate for  a  ``maximal multiple  point'', i.e.,
a multiple point in contact with a maximal number
of phases, may exist in terms of a set  of  coupling  parameters  that
comes about by  requiring  a  ``tightest  packing''  of  points  in  a
$N_{gen.}=3$ dimensional identification lattice embedded in a space with
a metric that is used to define the Manton action. At first sight,
this set of coupling parameters indeed appears to represent a point on
the boundary between a totally Coulomb phase and a ``large'' number of
phases that  are  confined  w.  r.  t.  various  invariant  subgroups.
However, upon  closer  inspection,  this  candidate  for  a  ``maximal
multiple point'' turns out to lie in a totally confined region of  the
action parameter space. This problem  may  however  be  resolvable  by
supplementing the Manton action with extra terms in such a way  as  to
relocate the phase boundaries. If this  is  feasible  and  if  such  a
``maximal multiple point'' is realized by Nature at the  Planck  scale
in the same way as we argue for the multiple point criticality for the
nonabelian groups, then, in the context of the anti-unification
scheme described in section 2,  the  diagonal  subgroup  inverse  fine
structure constant would be related to the  simple  critical  coupling
$\alpha_{U(1)crit}$ approximately through the relation

\begin{equation}
\frac{1}{\alpha_{diag}}=\frac{1}{\alpha_{crit}}
\left(N_{gen}+\left(\begin{array}{c}
N_{gen} \\ 2
\end{array} \right) \right).      \end{equation}

\nin Taking the number  of  $U(1)$  factors  to  be  $N_{gen.}=3$,
this   relation   yields   the   number   ``6''    for    the    ratio
$\alpha^{-1}_{diag}/\alpha^{-1}_{crit}$  which,  as  pointed  out  in
earlier work\cite{gosen}, is just what is needed to get agreement with
the Planck scale extrapolation\cite{kim,amaldi}  of  the  experimental
data for the $U(1)$ fine structure constant!

In connection with our ongoing endeavor to apply  the  multiple  point
criticality idea to the $U(1)$ part of the standard model symmetry, we
would like to mention what may be a noteworthy result. In  considering
the cartesian product of a large number  $L$  of  $U(1)$  factors,  it
appears that the maximization of the number of distinguishable  phases
that meet at the multiple point  (in  accord  with  the  principle  of
multiple point criticality) requires  a  plaquette  action  form  that
implements  a subdivision of the $L$  factors  of  $U(1)$
into ``bunches''of $U(1)$ factors with each bunch
containing  $N_{gen}=3$  $U(1)$
factors in the following sense: there  is  no
interaction between the $U(1)$ gauge  fields  from  different  bunches
whereas $U(1)$ factors within a bunch interact in a manner imposed  by
the assumption of a modified Manton action given by an  identification
lattice with hexagonal symmetry.

In an attempt to extend  the  idea
that many phases come together at the multiple point to  also  include
Higgs phases, it is speculated that this principle  might  suggest  a
system of Higgses in which each Higgs  field  only  couples  to  gauge
fields in one bunch. Within each  bunch,  the  principle  of  multiple
point criticality might  then  point  to  the  ``orthogonal''  arrangement
of  $N_{gen}$ Higgs
(meaning that these Higgs  fields  can  independently  be
Higgsed or not Higgsed). A Higgsing of a  subset  of  these  $N_{gen}$
Higgs could, for each bunch, subsequently  break  the  $SMG^{N_{gen}}$
gauge symmetry to the diagonal subgroup. If a remaining Higgs field happens
to have a Higgsed-unHiggsed phase transition that is almost second order, its
vacuum expectation value at the multiple point might be small enough so that
this Higgs field could be identified with the Weinberg-Salam Higgs.

With the above  mechanism, it can be
claimed that any large collection of $U(1)$ factors would be broken down  so
as to effectively show up as if there were only $N_{gen}$ $U(1)$ factors with
the rest  -  i.e.  the  $L/N_{gen}-1$  other  bunches  -  representing
decoupled matter (dark matter ?) reminiscent of the two  $E_8$  groups
in $SUSY$-string theories of the heterotic type. We hope to deal  with
these problems in connection with  our  ongoing  work  on  the  $U(1)$
coupling.

\section{Acknowledgment} It is with pleasure that we thank Christian Surlykke,
Svend Erik Rugh, Gudmar Thorleifsson and Larisa Laperashvili for useful and
inspiring discussions. Gudmar Thorleifsen is also thanked for the important
references \cite{bachas} and \cite{bhanot}. Support under European Economic
Community grant CS1-D 430-C is gratefully acknowledged.

\newpage

\addtolength{\topmargin}{-.25in}
\addtolength{\oddsidemargin}{-1.2in}
\addtolength{\evensidemargin}{-1.2in}

\nin \section*{Tables}

\nin Table 1 The quantity $Vol(H)$ for  any one of the 13
invariant subgroups $H$  of
the $SMG$ (these are listed in left  column),  can  be  written  as  a
product of some subset of the set of five  quantities  $(1/p_i)Vol(K_i)$.
The common set of factors  $1/p_{_{\tbz_2}}=1$,  $1/p_{_{\tbz_3}}=1$,
$1/p_{_{U(1)}}=1/6$,
$1/p_{_{SU(2)}}=1/2$, and $1/p_{_{SU(3)}}=1/3$, some subset of which  make
possible the factorization of all the $Vol(H)$  into  the  product  of
corresponding   subsets   of   the   quantities
$(1/p_{_{\tbz_2}})Vol(\bz_2)$,
$(1/p_{_{\tbz_3}})Vol(\bz_3)$,  $(1/p_{_{U(1)}})Vol(U(1))$,
$(1/p_{_{SU(2)}})Vol(SU(2))$,
and $(1/p_{_{SU(3)}})Vol(SU(3))$, are given in the last five columns.

\begin{small}

\[ \begin{array}{llc||ccccc}
Vol(H) & locally\;isomorph.\;cart.\;prod.\;gr. & \frac{1}{p_{_H}} &
\frac{1}{p_{_{\tbz_2}}} & \frac{1}{p_{_{\tbz_3}}}
& \frac{1}{p_{_{U(1)}}} & \frac{1}{p_{_{SU(2)}}} & \frac{1}{p_{_{SU(3)}}} \\ \\
Vol(\bunit) & & 1 &  &   &  &  &  \\
Vol(\bz_2) & Vol(\bz_2) & 1  & 1 &  &  &  &  \\
Vol(\bz_3) & Vol(\bz_3) & 1 &  & 1  & & &  \\
Vol(\bz_2 \times \bz_3) & Vol(\bz_2 \times \bz_3) & 1 & 1 & 1 &  &  &  \\
Vol(SU(2)) & Vol(SU(2)\times \bz_2) & 1/2 & 1 & &  & 1/2  &  \\
Vol(SU(3)) & Vol(SU(3)\times \bz_3) & 1/3 & & 1 & &  & 1/3 \\
Vol(SU(2)\times  \bz_3) & Vol(SU(2)\times \bz_2\times  \bz_3) & 1/2 & 1 & 1 &
& 1/2 &  \\
Vol(SU(3)\times \bz_2) & Vol(SU(3)\times \bz_3 \times \bz_2) & 1/3 & 1 & 1 &
& & 1/3  \\
Vol(U(1)) & Vol(U(1)\times \bz_2\times \bz_3) & 1/6 & 1 & 1 & 1/6 &  & \\
Vol(SU(2)\times SU(3)) & Vol(SU(2)\times SU(3)\times \bz_2\times \bz_3) & 1/6 &
1 & 1 &  & 1/2 & 1/3  \\
Vol(U(3)) & Vol(U(1)\times SU(3)\times \bz_2\times \bz_3) & 1/18 & 1 & 1 & 1/6
& & 1/3 \\
Vol(U(2)) & Vol(U(1)\times SU(2)\times \bz_2\times \bz_3) & 1/12 & 1 & 1 & 1/6
& 1/2 &  \\
Vol(SMG) & Vol(U(1)\times SU(2)\times SU(3)\times \bz_2\times \bz_3) & 1/36  &
1 & 1 & 1/6 & 1/2 & 1/3
\end{array} \]

\end{small}

\vspace{.5cm}

\nin Table 2 In a  manner  analogous  to  that
of  Table~1, the quantities $Vol(H)$ for the  5
invariant subgroups $H$ of $U(N)$ (listed in left column)  factorize  into
products of subsets of the constituent  quantities  $(1/p_i)Vol(K_i)$.
The coefficient $1/p_i$ of any corresponding $Vol(K_i)$ is, as seen in
the last three columns, the same for all the $Vol(H)$ in which such  a
$Vol(K_i)$ contributes in the factorization of $Vol(H)$.
Figures 1 and 2, which depict the phase diagram for $U(2)$, illustrate
how the 5 \pcps of a $U(N)$ group meet at the multiple point in our
approximation.

\[ \begin{array}{llc||ccc}
Vol(H)& locally\;isomorphic\; cart.\;prod.\;gr. & \frac{1}{p_{_H}} &
\frac{1}{p_{_{\tbz_N}}} & \frac{1}{p_{_{U(1)}}} & \frac{1}{p_{_{SU(N)}}} \\ \\
Vol(\bunit) & & 1 &  & & \\
Vol(\bz_N) & Vol(\bz_N) & 1 & 1 & & \\
Vol(U(1)) & Vol(U(1) \times \bz_N)  & 1/N & 1 & 1/N &  \\
Vol(SU(N)) & Vol(SU(N)\times \bz_N) & 1/N & 1 &  & 1/N \\
Vol(U(N)) & Vol(U(1)\times SU(N)\times \bz_N) & 1/N^2 & 1 & 1/N & 1/N

\end{array} \]

\addtolength{\oddsidemargin}{1.2in}
\addtolength{\evensidemargin}{1.2in}

\begin{small}
\begin{tabular}{|p{2.50 in}|l|}
\hline
\multicolumn{2}{|c|}{Table 3: SU(2) Gauge Coupling} \\
\hline \hline
\raggedright
Prediction~for~continuum~limit coupling~estimate,~$1/\alpha_
{2,\; triple\;point,\;cont.}$, using  & \\
\raggedright {\footnotesize 1.~not~exponentiated:} & $\overbrace{0.71\cdot   20
  }^{14.2}+\overbrace{0.89\cdot
1.7}^{1.5}=15._7    \pm    1$    \\     \raggedright     {\footnotesize
2.~exponentiated:}               &               $\overbrace{0.75\cdot
20}^{15._0}+\overbrace{0.89\cdot  1.7}^{1.5}=16._5  \pm  1$  \\  \hline
Experimental value\cite{kim,amaldi} for & \\ $1/\alpha_2$ reduced by a factor
3: & $\frac{1}{3}\cdot \alpha^{-1}_2(M_Z)=  \frac{1}{3}\cdot  (29.7\pm
0.2)=9.9\pm 0.07$ \\ ``desert extrapolation\cite{kim,amaldi}''  to  &
\\ Planck scale with one Higgs: &  $\stackrel{desert}{\longrightarrow}
\frac{1}{3}\cdot \alpha^{-1}_2 (\mu_{Pl.})=\frac{1}{3}\cdot 49._{5}=16.5$
\\ & \\ \hline \hline $\beta_{adj.,\;triple\;point}$ {\footnotesize  (i.e.,
at triple point)} & 2.4 (\mbox{ca. 5\% uncertainty from  MC})  \\
\hline  $\beta_{f\;triple\;point}$  {\footnotesize  (i.e.,  at  triple
point)} &  0.54  (\mbox{ca.  10\%  uncertainty  from  MC})  \\  \hline
\raggedright  $\beta_{adj.}$-contribution~to~$1/\alpha_{2,\;triple\;point}$
&  \\  {\footnotesize  (without   continuum   correction)}   &   $4\pi
\frac{C^{(2)}_{adj.}}{(2^2-1)}\beta_{adj,\;triple\;point}=4\pi \cdot  (2/3)
\cdot 2.4=20$ \\ & \\  \hline  \raggedright  $\beta_f$-contribution~to
{}~$1/\alpha_{2,\;triple\;point}$         &         \\         {\footnotesize
(without~continuum~correction):}                &                $4\pi
\frac{C^{(2)}_f}{(2^2-1)}\beta_{f\;triple\;point}=4\pi                \cdot
(\frac{3}{4}/3)\cdot  0.54=1.7$  \\  &  \\  \hline  \raggedright  Full
$1/\alpha_{2,\;triple\;point}$  {\footnotesize  (without  continuum}  &  \\
{\footnotesize                     correction):}                     &
$1/\alpha_{2,\;triple\;point,\;full,\;no\;cont.}=20+1.7=21.7$   \\   \hline
\raggedright       Continuum       correction        factor        for
$\beta_{adj.}$-contribution:  &  \\  \raggedright  {\footnotesize   1.
not~exponentiated (using (\ref{corbian})):}  &  $1-C^{(2)}_{adj.}  \pi
\alpha_{2,\;triple\;point,\;full,\;no.\;cont.}=$            \\            &
$1-2\pi/21.7=1-0.290=0.71$   \\   \raggedright    {\footnotesize    2.
exponentiated:}          &          $\exp(-C^{(2)}_{adj.}          \pi
\alpha_{2,\;triple\;point,\;full,\;no.\;cont.})$=           \\            &
$\exp(-2\pi/21.7)=\exp(-0.290)=0.75$  \\  &  \\  \hline   \raggedright
Continuum  correction  factor   for   $\beta_f$-contribution:   &   \\
\raggedright    {\footnotesize     1.     not~exponentiated     (using
(\ref{corbian})):}            &            $1-C^{(2)}_f            \pi
\alpha_{2,\;triple\;point,\;full,\;no.\;cont.}=$  \\   &   $1-(\frac{3}{4})
\pi/21.7=1-0.109=0.89$    \\    \raggedright     {\footnotesize     2.
exponentiated:}           &            $\exp(-C^{(2)}_f            \pi
\alpha_{2,\;triple\;point,\;full,\;no.\;cont.})=$ \\ & $\exp(-(\frac{3}{4})
\pi/21.7)=\exp(-0.109)=0.90$ \\ & \\ \hline \end{tabular} \end{small}


\begin{small}
\begin{tabular}{|p{2.50 in}|l|}
\hline
\multicolumn{2}{|c|}{Table 4: SU(3) Gauge Coupling} \\
\hline \hline
\raggedright
Prediction~for~continuum~limit coupling~estimate,~$1/\alpha_{3,\;
triple\;point,\;cont.}$, using  & \\
\raggedright {\footnotesize 1.~not~exponentiated:} & $\overbrace{0.65\cdot
25}^{16._3}+\overbrace{0.84\cdot
1.7}^{1.4}=17._7    \pm    1$    \\     \raggedright     {\footnotesize
2.~exponentiated:}               &               $\overbrace{0.70\cdot
25}^{17._5}+\overbrace{0.85\cdot  1.7}^{1.4}=18._9  \pm  1$  \\  \hline
Experimental value\cite{kim,amaldi} for & \\ $1/\alpha_3$ reduced by a factor
3:   &    $\frac{1}{3}\cdot\alpha^{-1}_3(M_Z)=\frac{1}{3}\cdot(8.47\pm
0.5)=2.8\pm 0.2$ \\ ``desert extrapolation\cite{kim,amaldi}'' to & \\
Planck scale with  one  Higgs:  &  $\stackrel{desert}{\longrightarrow}
\frac{1}{3}\cdot \alpha^{-1}_3(\mu_{_{Pl.}})=\frac{1}{3}\cdot  53  \pm
0.7 =17.7\pm 0.3$ \\  &  \\  \hline  \hline  $\beta_{adj.,\;triple\;point}$
{\footnotesize  (i.e.,  at  triple  point)}  &   5.4   (ca.   5\%
uncertainty) \\ \hline $\beta_{f\;triple\;point}$ {\footnotesize (i.e.,  at
triple  point)}  &  0.8  (ca.   20\%   uncertainty)   \\   \hline
\raggedright  $\beta_{adj.}$-contribution~to~$1/\alpha_{3,\;triple\;point}$
&  \\  {\footnotesize  (without   continuum   correction)}   &   $4\pi
\frac{C^{(2)}_{adj.}}{(3^2-1)}\beta_{adj,\;triple\;point}=4\pi \cdot  (3/8)
\cdot 5.4=25$ \\ & \\  \hline  \raggedright  $\beta_f$-contribution~to
{}~$1/\alpha_{3,\;triple\;point}$         &         \\         {\footnotesize
(without~continuum~correction):}                &                $4\pi
\frac{C^{(2)}_f}{(3^2-1)}\beta_{f\;triple\;point}=4\pi                \cdot
(\frac{4}{3}/8)\cdot  0.8=1.7$  \\  &  \\  \hline  \raggedright   Full
$1/\alpha_{3,\;triple\;point}$  {\footnotesize  (without  continuum}  &  \\
{\footnotesize                     correction):}                     &
$1/\alpha_{3,\;triple\;point,\;full,\;no\;cont.}=25+1.7=26._7$  \\   \hline
\raggedright       Continuum       correction        factor        for
$\beta_{adj.}$-contribution:  &  \\  \raggedright  {\footnotesize   1.
not~exponentiated (using (\ref{corbian})):}  &  $1-C^{(2)}_{adj.}  \pi
\alpha_{3,\;triple\;point,\;full,\;no.\;cont.}=$            \\            &
$1-3\pi/26.7=1-0.35=0.65$   \\    \raggedright    {\footnotesize    2.
exponentiated:}          &          $\exp(-C^{(2)}_{adj.}          \pi
\alpha_{3,\;triple\;point,\;full,\;no.\;cont.})$=           \\            &
$\exp(-3\pi/26.7)=\exp(-0.35)=0.70$  \\  &  \\   \hline   \raggedright
Continuum  correction  factor   for   $\beta_f$-contribution:   &   \\
\raggedright    {\footnotesize     1.     not~exponentiated     (using
(\ref{corbian})):}            &            $1-C^{(2)}_f            \pi
\alpha_{3,\;triple\;point,\;full,\;no.\;cont.}=$  \\   &   $1-(\frac{4}{3})
\pi/26.7=1-0.16=0.84$    \\     \raggedright     {\footnotesize     2.
exponentiated:}           &            $\exp(-C^{(2)}_f            \pi
\alpha_{3,\;triple\;point,\;full,\;no.\;cont.})=$ \\ & $\exp(-(\frac{4}{3})
\pi/26.7)=\exp(-0.16)=0.85$ \\ & \\ \hline \end{tabular} \end{small}

\newpage

\nin \section*{Figure Captions}

\begin{itemize}

\item [{Fig. 1}] The region of allowed parameters
($\log Vol(SU(2)),\log Vol(U(1)),
\log Vol(\bz_2)$) for the modified Manton action: $\log(\pi e)^{3/2}\leq
\log Vol(SU(2)) \;(\approx \frac{3}{2}\log\beta_2+\log vol(SU(2)))< \infty,
\log (\pi e)^{1/2}\leq
\log Vol(U(1)) \;(\approx \frac{1}{2}\log\beta_1+\log vol(U(1)))< \infty,
0\leq \log Vol(\bz_2) \leq \log 2$. These intervals reflect our having
used $Vol H$ that, up to a factor $(\pi e)^{\frac{dim(H)}{2}}$,
are measured in units of the fluctuation volume. The cube with the chopped off
corner
represents the region of total confinement. Walls that extend to
$+\infty$ are terminated in the drawing with
irregular wavy boundaries.

\item [{Fig. 2}] Phase diagram for lattice gauge theory
with gauge group $U(2)$ in our weak
coupling approximation with modified Manton action. We have drawn the
figure with positive  effective
dimension for the discrete constituent invariant
subgroup $\bz_2$.
Rectangular signs on signposts are marked
with the confining
invariant subgroup $H$ and indicate the regions corresponding
to the 5 possible
phases; these 5 phases are seen to meet at the multiple point.
The oval signs lie in the phase boundaries and specify the factor group
$L=H_1/H_2$ formed from the two invariant subgroups $H_1$ and $H_2$
that are confined on the two sides of the boundary. It is these groups $L$
that change behavior in crossing the phase boundary in question.
Unbroken shading lines indicate phase boundaries as seen from within the
totally Coulomb-like phase.

\end{itemize}


\newpage

\begin{thebibliography}{99}

\bibitem{mich} L. Michel, Invariance in Quantum Mechanics and Group Extension,
appearing in Group Theoretical Concepts and Methods in Elementary Particle
Physics,  Lectures of the Istanbul Summer School of Theoretical Physics,
July 16 - August 4, 1962 (edited by F. G\"{u}rsey, Gorden and Breach Science
Publishers).


\bibitem{group}
L. O' Raifeartaigh, Group Structure of Gauge Theories, Cambridge Univ. Press,
Cambridge, 1986.

\bibitem{confusion}
H.B. Nielsen and N. Brene, Gauge Glass, Proc.  of  the  XVIII
International  Symposium  on  the  Theory  of   Elementary   Particles,
Ahrenshoop, 1985 (Institut   fur   Hochenergiphysik,   Akad.   der
Wissenschaften der DDR, Berlin-Zeuthen, 1985);

D.L. Bennett, N. Brene, L. Mizrachi and H.B.  Nielsen,  Phys.  Lett.
{\bf B178} (1986) 179.

\bibitem{gaeta}
G. Gaeta, Int. J. Theor. Phys. {\bf 32} (1993) 727

\bibitem{cop} H. Flyvbjerg, B. Lautrup and J.B. Zuber, Phys.Lett {\bf 110B}
(1982) 279;

J. Greensite and B. Lautrup, Phys. Lett {\bf 104B} (1981) 41;

P. Cvitanovi\'{c}, J. Greensite and B. Lautrup, Phys. Lett {\bf 105B} (1981)
197;

J. Greensite and B. Lautrup, Phys Rev. Lett. {\bf 47} (1981) 9;

B. Lautrup and M. Nauenberg, Phys. Rev Lett. {\bf 45} (1980) 1755.

\bibitem{randyn}  H.B.  Nielsen,  Dual  Strings (see   Section
``Catastrophe Theory''), in Fundamentals of Quark Models,  Proc.  of
the XVII Scottish University Summer School  in  Physics,  St  Andrews,
1976 (eds. I.M. Barbour and A.T. Davies,  University  of  Glasgow,
Scotland, 1976, p. 528;

H.B. Nielsen, D.L. Bennett and N. Brene, in Recent Developments in
Quantum Field Theory, Proc. of  the  Niels  Bohr  Centennial  Conf.,
Copenhagen, 1985 (eds.  J. Ambj\o rn, B.J.  Durhuus  and  J.L.  Petersen,
North Holland, Amsterdam, 1985);

D.L. Bennett, N. Brene, and H.B. Nielsen, Random  Dynamics,
Nobel  Symposium:  Unification  of  Fundamental   Interactions
(World  Scientific, London, 1988; copublished in Physica Scripta {\bf T15}
(1987) 158)

C.D. Froggatt and H.B.  Nielsen,  Origin  of  Symmetries,  World
Science Pub. Co., Inc., Singapore, 1991.

\bibitem{fnns} D. F\"{o}rster, H. B. Nielsen and M. Ninomiya: Phys. Lett
{\bf 94B} (1980) 135

S. Shenker, Application to DOE (1980, unpublished)

H.B. Nielsen and N. Brene, ``Spontaneous emergence of gauge symmetry'',
in Proc. of the Workshop on Skyrmions and Anomalies, M. Jezabek and M.
Praszalowicz (Eds.), World Science Publishing, Singapore, 1987, p. 493

C.D. Froggatt  and  H.B.  Nielsen,  Origin  of
Symmetries, World Science Pub. Co., Inc., Singapore (1991) p. 110

\bibitem{skewness} H.B.  Nielsen  and  N.  Brene,  Skewness  of  the
Standard  Model:  Possible  Implications,  Physicalia  Magazine,   The
Gardener of Eden, {\bf 12} (1990) 157;

H.B. Nielsen and N. Brene, Phys. Lett. {\bf223} (1989) 399.

\bibitem{anomalies}  S.  Chadha  and  H.   B.   Nielsen   and   Chadha
(unpublished), but see page 26 at the position of reference number 10
in C. Froggatt and H. B.
Nielsen ``Origins  of  Symmetries'',  World  Science  Pub.  Co.,  Inc.,
Singapore (1991)

C. Q. Geng and R. E. Marshak, Phys. Rev. D {\bf 39}, 693,  (1989)  and
Phys. Rev. D {\bf 41}, 717, (1990)

\bibitem{korfu} H.B. Nielsen, S.E. Rugh and C. Surlykke, Seeking Inspiration
from  the
Standard Model in Order to Go Beyond It, Proc. of Conference  held  on
Korfu (1992)

\bibitem{gosen} H.B. Nielsen and  D.L.  Bennett,  Fitting  the  Fine
Structure Constants by Critical Couplings and Integers, Proc.  of  the
XXV International Ahrenshoop Symposium on  the  Theory  of  Elementary
Particles, Gosen, Sept. 23 - 26, 1991.

\bibitem{van}  D.L.Bennett and H.B. Nielsen, Standard Model Couplings from
Mean Field Criticality at the Planck Scale and a Maximum Entropy Principle,
Proc. of the Vancouver Meeting on Particles and Fields '91 (eds. D. Axen,
D. Bryman, and M. Comyn, World Scientific, Singapore, 1992, p. 857).

\bibitem{long} D.L. Bennett and H.B. Nielsen, The  Three  Yang-Mills
Coupling Constants from Planck Scale Criticality and a Maximum Entropy
Principle, NBI-HE-92-02.

\bibitem{lap} L.V. Laperashvili, The Standard Model and Fine Structure
Constants on Planck Scale in Random Dynamics of Bennett-Brene-Nielsen-Picek
(half review in Russian), submitted to Yadernaya Fizika (1992)

\bibitem{ngen} H.B. Nielsen, Acta Phys. Pol. {\bf B20} (1989) 427;

H.B. Nielsen, The Number of Fermion Generations and the
Fine Structure Constants, Proc. of the XXVIII Cracow School of Theor.
Phys., Zakopane, Poland, 10-20 may 1988. (Random Dynamics and Relations
Between the Number of Fermion Generations and the Fine Structure Constants,
NBI-HE-89-01 (preprint only));

H.B. Nielsen, On the Number of Generations in the Standard Model,
Proc. of the Int. Sem. {\em Quarks '88,} Tblilisi, USSR, 17-21 May 1988
(eds. A. N. Tavkhelidze et al., World Scientific, Singapore, 1989, p. 81-89);

D.L. Bennett, H.B. Nielsen  and  I.  Picek,  Number  of  Generations
Related to Coupling Constants by  Confusion,  Proc.  XX  International
Symposium  on  the  Theory  of  Elementary  Particles, Ahrenshoop,
October 13-17, 1986 (Akad.  der Wissenschaften der DDR, Berlin  p. 383);

H.B.  Nielsen,  D.L. Bennett  and  I.  Picek,
Formulation  Using  Compactified  Dimensions  to  get  an   Inequality
Relating Coupling Constants and the Number of  Generations,   DST
Workshop on Particle Physics  -  Superstring  Theory,  Kanpur,  India
13-24 Dec. 1987 (World Scientific, Singapore, 1988, p. 324-46);

H.B. Nielsen, Random Dynamics, Three Generations, and Skewness,
Proc. of Summer Meeting on Quantum Mechanics of Fundamental Systems III,
Santiago, Chile, Jan. 9-12, 1990 (also preprint NBI-HE-91-04);

D.L. Bennett, H.B. Nielsen and I. Picek, Understanding
Fine Structure Constants and Three Generations, Phys. Lett. {\bf 208B}
(1988) 275;

D.L. Bennett, H.B. Nielsen and I. Picek,
Understanding Fine Structure  Constants  and  Three  Generations,
Proc. of the XI Workshop on High Energy  Physics  and  Field  Theory,
Protvino, U. S. S. R. July 5-9, 1988 (Academy of  Sciences  of
the U. S. S. R., Moscow, 1989, p. 419-27);

D.L. Bennett, H.B. Nielsen and I. Picek, Gauge  Group
Replication as an Explanation for the Smallness of the Fine  Structure
Constant, Proc. of the XXI International Symposium on the  Theory
of Elementary  Particles,  Sellin,  Oct  12-16,  1987  (Institut  fur
Hochenergie   Physik,   Akad. der   Wissenschaften    der    DDR,
Berlin-Zeuthen, 1987, p. 287-308);

H.B. Nielsen, D.L. Bennett and I. Picek, An  Inequality
Relating Gauge Group Coupling Constants and the Number of Generations,
Proc. of the Beijing Workshop on String Theories, Beijing, July,
1987 (World Scientific, Singapore, 1988);

H.B. Nielsen and N. Brene, Some Remarks on Random Dynamics, Proc.
of the 2nd Nishinomiya Yukawa Memorial Symposium on String Theory, Kyoto
University, 1987 (Springer, Berlin, 1988);





\bibitem{kim} H.B. Kim and J.E. Kim, Coupling  Constant  Unification
and LEP Data, Preprint SNUTP 92-89.

\bibitem{amaldi}
U. Amaldi et al, CERN-PPE/91-44, 22 March 1991;

U. Amaldi et al, Phys.Rev. {\bf D36} (1987) 1385;

J. Ellis et al, Phys.Lett. {\bf B249} (1990) 441 and CERN-TH-5943/90;

P. Langacker, Univ. of Penn. preprint UPR-0435T (1990);

DELPHI collaboration: P. Abreu et al, Phys.  Lett.  {\bf  247B}
(1990) 167 and Phys. Lett. {\bf 252B} (1990) 149.





\bibitem{gap}  C.D.  Froggatt,  G. Lowe,  and  H.B.   Nielsen,
The Fermion Hierarchy and Gauged Chiral Flavour  Quantum Numbers,
Glasgow University Preprint (1993).

C.D. Froggatt, G.  Lowe  and  H.B.  Nielsen,  Anti-grand
Unification and the Fermion Mass Problem, Glasgow University  Preprint
GUTPA-93-07-01 (1993)

C.D. Froggatt, G. Lowe  and  H.B.  Nielsen,  Fermion  Masses  and  the
Standard Model Group as a  Diagonal  Subgroup,  Niels  Bohr  Institute
Preprint (1993)

\bibitem{frogniel} C.D. Froggatt  and  H.B.  Nielsen,  Origin  of
Symmetries, World Science Pub. Co., Inc., Singapore (1991) p. 116.

\bibitem{bachas} C.P. Bachas and R.F.  Dashen, Nucl.  Phys  {\bf  B210}
(1982) 583.

\bibitem{bhanot} G. Bhanot, Phys. Lett. {\bf 108B} (1982) 337.

\bibitem{drouffe} J.-M. Drouffe and J.-B. Zuber, Phys. Rep. {\bf 102} (1983)
1.




\bibitem{cen} H. B. Nielsen and N. Brene, Nucl. Phys. {\bf B224} (1983) 396;

Gauge Glass Model, Proc. of Division of Particles and Fields of APS,
Santa Fe, Oct. 31 - Nov. 3, 1984 (eds T. Goldman and M.M. Nieto, World
Scientific, 1984, p. 352;

H.B. Nielsen and D.L. Bennett, The Gauge Glass: a short review,
Nordita preprint 85/23 (1985).

\bibitem{baby} H.B. Nielsen and M.N. Ninomiya, Baby Universes,  Fine
Tuning Problems - A Theory of Everything  Robbing the Throne by Killing
the Rivals, Proc. of the XXII International Symposium on the Theory of
Elementary Particles, Ahrenshoop, DDR, Oct. 17-21, 1988.

\bibitem{sur} C. Surlykke, On Monopole Suppression in Lattice QED Gauge
Theories, in preparation, soon to appear as NBI Preprint.

\bibitem{mit} V.G. Bornyakov, V.K. Mitrjushkin and M. M\"{u}ller-Preussker,
Nucl. Phys. (Proc. Suppl.) {\bf 30} (1993) 587.

\bibitem{cs} Christian Surlykke, private communication.

\bibitem{bjones} Bernard Jones, private communication.

\bibitem{strings}  G. 't Hooft, Nucl. Phys. {\bf 138} (1978) 1

H.B. Nielsen, ``Confinement with emphasis on the Copenhagen Vacuum'' In:
Particle Physics 1980, Proc. of the 3rd Adriatic Summer Meeting,
Dubrovnik, Yugoslavia; I. Andric, I Dadic and N. Zovko (Eds.), North-Holland
Pub. Co., 1981, p. 67.

\end{thebibliography}
\end{document}